\documentclass[smallextended]{svjour3}

\smartqed

\usepackage{url}
\usepackage{textcomp}
\usepackage{balance}
\usepackage[most]{tcolorbox}
\usepackage{enumitem}
\usepackage{caption}
\usepackage{subcaption}
\usepackage{threeparttable}
\usepackage{multirow}
\usepackage{makecell}
\usepackage{wrapfig}
\newlist{indenteddesc}{description}{1}
\setlist[indenteddesc,1]{leftmargin=3em, labelindent=1em}
\AtBeginDocument{%
  }

\newcommand\tool[1]{LLMAID}

\newcommand\etal[1]{\textit{et al.}}
\newcommand\ie[1]{\textit{i.e.}}
\newcommand\eg[1]{\textit{e.g.}}

\begin{document}

\title{\tool{}: Identifying AI Capabilities in Android Apps with LLMs}

\author{Pei Liu* \thanks{*This work was mostly done when Pei Liu was a Research Fellow in CSIRO's Data61}  \and
        Terry Yue Zhuo \and
        Jiawei Deng \and
        James Thong \and
        Shidong Pan \and
        Sherry (Xiwei) Xu \and
        Zhenchang Xing \and
        Qinghua Lu \and
        Xiaoning Du \and
        Hongyu Zhang
}

\institute{Pei Liu \at
              Chongqing University, China \\
              \email{peiliu@cqu.edu.cn}           %  \\
           \and
            Terry Yue Zhuo, Xiaoning Du \at
              Monash University, Australia \\
              \email{\{terry.zhuo, xiaoning.du\}@monash.edu}  
           \and
            Jiawei Deng \at
              CSIRO's Data61, Australia \\
              \email{dengjiaweibryan@gmail.com}
           \and
            James Thong, Sherry (Xiwei) Xu, Zhenchang Xing, Qinghua Lu \at
              CSIRO's Data61, Australia \\
              \email{\{james.hoang, xiwei.xu, zhenchang.xing, qinghua.lu\}@data61.csiro.au}
           \and
           Shidong Pan \at
            New York University \& Columbia University, USA \\
            \email{Shidong.Pan@anu.edu.au}
           % \and
           % Sherry (Xiwei) Xu \at
           %  CSIRO's Data61, Australia \\
           %  \email{xiwei.xu@data61.csiro.au}
           % \and
           %  Zhenchang Xing \at
           %    CSIRO's Data61, Australia \\
           %    \email{zhenchang.xing@data61.csiro.au}  
           %  \and
           %  Qinghua Lu \at
           %    CSIRO's Data61, Australia \\
           %    \email{qinghua.lu@data61.csiro.au}  
           %  \and
           %  Xiaoning Du \at
           %    Monash University, Australia \\
           %    \email{xiaoning.du@monash.edu}  
            \and
            Hongyu Zhang \at
              Chongqing University, China \\
              \email{hyzhang@cqu.edu.cn} 
}

\date{Received: date / Accepted: date}
\maketitle

\begin{abstract}
% The advances in Artificial Intelligence (AI) and its applications in mobile devices have gained significant attention from researchers to investigate the characterizations of AI capabilities in mobile applications (apps). 
% However, existing works solely employ manual inspections and unscalable rule-based approaches, which are costly and cannot be generalized to emerging AI techniques. In addition, these methods only focus on specific types of information like AI package, APIs and on-device AI models, and ignore other potential AI candidates like file names and HTTPS requests. Thus, they may not provide a comprehensive overview of all AI services provided in apps. 
Recent advancements in artificial intelligence (AI) and its widespread integration into mobile software applications have received significant attention, highlighting the growing prominence of AI capabilities in modern software systems. However, the inherent hallucination and reliability issues of AI continue to raise persistent concerns. Consequently, application users and regulators increasingly ask critical questions such as: \textit{``Does the application incorporate AI capabilities?''} and \textit{``What specific types of AI functionalities are embedded?''}
% Application consumers and regulators are more and more curious about, \textit{Does the app integrate AI capabilities?} and even \textit{What types of AI capabilities have been integrated?} 
Preliminary efforts have been made to identify AI capabilities in mobile software; however, existing approaches mainly rely on manual inspection and rule-based heuristics. These methods are not only costly and time-consuming but also struggle to adapt advanced AI techniques. Moreover, they lack the granularity needed to accurately characterize the diverse range of AI functionalities.
% Although some efforts have been made to uncover AI capabilities in mobile software artifacts, existing approaches rely mainly on manual inspections and rule-based methods, which are costly and ineffective in adapting to emerging AI techniques, and also agnostic to fine-grained functional characteristics of AI capabilities. 
Specifically, existing methods often focus on specific types of information, such as AI packages and on-device models, while overlooking other potential AI factors, \ie{}, APIs and HTTPS requests. As a result, these methods fail to provide a comprehensive AI service summarization of software applications.

To address the limitations of existing methods, we propose \tool{} (\textit{Large Language Model for AI Discovery}), a semantics-driven framework that combines program analysis techniques with large language models (LLMs) to identify and analyze AI services in mobile applications. \tool{} includes four main tasks: (1) candidate extraction, (2) knowledge base interaction, (3) AI capability analysis and detection, and (4) AI service summarization. 
% a \textit{\underline{LLM} for \underline{AI} \underline{D}iscovery} (\tool{}) framework, a semantic-based approach that integrates program analysis techniques with LLMs to identify and analyze AI services in apps. \tool{} includes four main tasks: candidate extraction, AI candidate determination, AI capability analysis and detection, and AI service summarization. 
We apply \tool{} to a dataset of 4,201 Android applications and demonstrate that it identifies 242\% more real-world AI apps than state-of-the-art rule-based approaches. Our experiments show that LLMAID achieves high precision and recall, both exceeding 90\%, in detecting AI-related components. Additionally, a user study indicates that developers find the AI service summaries generated by \tool{} to be more informative and preferable to the original app descriptions.
% We analyze 4,201 Android apps and \tool{} detects 242\% more real-world AI apps compared to the state-of-the-art rule-based approaches. 
% Our experimental results demonstrate high precision and recall (over 90\%) in identifying AI components, and our user study reveals that developers prefer AI summaries generated by \tool{} over original app descriptions. 
Finally, we leverage \tool{} to perform an empirical analysis of AI capabilities across Android apps. The results reveal a strong concentration of AI functionality in computer vision (54.80\%), with object detection emerging as the most common task (25.19\%).
\end{abstract}


\section{Introduction}

% ~\cite{namysl2019efficient,yin2019deep}, ~\cite{gupta2021deep,bhalla2020simulation,rao2018deep}, ~\cite{bovet2000picture,bansal20212d,zou2023object}, ~\cite{rosa2018knowledge,sahoo2019deepreco,zhang2019deep}, tflite:2023,tensor:2023,
Advancements in artificial intelligence (AI) and their successful integration into mobile applications (apps) have enhanced the capabilities of Android apps in daily activities, such as optical character recognition, autonomous driving, object recognition and shopping recommendations.
To support developers' integration of AI capabilities, several vendors have introduced mobile AI frameworks, such as TensorFlow Lite~\cite{abadi2016tensorflow}, Caffe2~\cite{caffe2:2023}, Core ML~\cite{coreml:2023}, and ncnn~\cite{ncnn:2023}. 
These frameworks enable efficient on-device model building and inference, offering better privacy protection and reducing bandwidth usage compared to cloud-based models~\cite{guo2018cloud}. 
% These frameworks aim to provide timely on-device model building and inference. Moreover,  Compared to the cloud-based models~\cite{guo2018cloud}, on-device model inference better protects user privacy and saves bandwidth load. 
In recent years, there has been a surge in the development of large AI models, such as stable diffusion~\cite{stable:2023}, ChatGPT~\cite{chatgpt:2023}, LLaMA~\cite{touvron2023llama}, and DeepSeek~\cite{liu2024deepseek}. With their advanced abilities to understand and generate content, emerging app developers increasingly resort to these models via the cloud to enhance their businesses~\cite{xu2019first,hu2023first}. 

With the widespread adoption of AI and the growing awareness of its inherent limitations, issues of AI discoverability and transparency have attracted attention from both application consumers and various regulatory bodies. 
To have a comprehensive understanding of AI services in software artifacts for these stakeholders, this study first tries to uncover the current status quo of AI services in real-world Android applications. 
In this paper, we define this process as \textbf{AI Discovery}, which encompasses determining whether AI capabilities are integrated or not and further elucidating the functional characteristics of the incorporated AI services.  
Several studies have been conducted to investigate AI techniques in smartphone apps~\cite{xu2019first,zhao2022survey,li2022ai,siu2023towards,deng2019deep}. % wang2022survey, cai2022enable, 
However, these studies exhibit several notable constraints: 
\textit{1)} %Prior work focuses on identifying the characteristics of on-device AI models, 
\textbf{lacking attention} to other potential AI components,\footnote{Component in the paper may refer to either high-level modules or low-level elements such as APIs and libraries, depending on context.} such as integrated development packages, internal assets, and HTTP requests and API invocations in detailed implementations and integrated libraries. Ignoring these components would lead to an incomplete and insufficient analysis; 
\textit{2)} relying on \textbf{manual inspection} or \textbf{pre-defined keyword} searches, which require up-to-date human expertise, making them expensive and difficult to scale; 
\textit{3)} depending on human-defined \textbf{rule-based} methods, which are constrained by domain knowledge and thus ineffective for identifying AI components in diverse real-world applications. 

To address these aforementioned challenges, we argue that automating the human-like \textbf{semantic-based AI Discovery} is essential, enabling the characterization of real-world AI components by interpreting their meaningful names. 
Inspired by recent findings~\cite{hagendorff2023human,ning2024can,li2024llms,pan2024large} that large language models (LLMs) are capable of exhibiting human-like reasoning and decision-making, we explore how these models can enhance the AI Discovery process by interpreting semantically meaningful AI component names.  % jiang2023knod, nijkamp2022conversational, dakhel2023github,
Moreover, recent studies have shown that these LLMs can automate various software tasks, including automatic program repair~\cite{zhang2022repairing,jiang2023impact,xia2023conversational,han2024chase}, code generation~\cite{nijkamp2022conversational,chakraborty2022natgen,fried2022incoder,liao20243}, and improving software analysis efficiency and effectiveness~\cite{ma2023scope,li2023hitchhiker}. 
These findings provide evidence that LLMs may have substantial knowledge in software practice, motivating us to explore their full potential in AI Discovery within mobile applications.

To this end, we propose \tool{} (\textit{Large Language Model for AI Discovery}), the first framework that leverages large language models (LLMs) to identify AI components extracted through static program analysis and interpret their functional capabilities. 
% To enable LLMs to handle these complex tasks effectively, we adopt the \textit{AI Chains} technique~\cite{wu2022ai} to guide the prompt design in tasks involving LLM interaction, which enhances automation while maintaining both reliability and transparency. 
The \tool{} framework consists of four key stages: (1) candidate extraction, (2) knowledge base interaction, (3) AI capability analysis and detection, and (4) AI service summarization. This structured approach allows us to perform AI discovery in an automated, step-by-step manner, significantly reducing the need for manual intervention.

We implemented \tool{} using the widely adopted GPT-3.5-turbo model, which is much more cost-effective than the advanced GPT-4 model~\cite{gptmodel:2023}. 
To evaluate the effectiveness of \tool{}, we conducted validation on a sample of 382 AI applications selected from a total of 56,682 already identified AI apps~\cite{li2022ai}. 
Manual evaluation of the results confirms that \tool{} achieves high precision and recall (exceeding 90\%) in detecting AI components.
Furthermore, we applied \tool{} to a dataset of 3,819 real-world Android apps published on Google Play between 2019 and 2023. 
Compared to the state-of-the-art (SOTA) detection approach, AI Discriminator~\cite{li2022ai},  \tool{} identifies more than twice as many AI Android apps, showcasing its superior detection capability. 
Our analysis reveals that the majority of detected AI components pertain to the domains of computer vision (54.80\%) and data analysis (26.85\%), with object detection (25.19\%) and data processing (23.17\%) being the most prevalent tasks. 
At the application level, we find that 47.44\% (643/1366) of AI-enabled apps offer services related to data analysis, while 46.41\% (634/1366) incorporate computer vision functionalities. 
% We further validate the practical value of \tool{} by showing that developers find its AI service reports informative and preferable. 
We further demonstrate the practical utility of~\tool{} by evidencing that developers perceive its AI service report as both informative and preferable. 

Our main contributions are as follows:
\begin{itemize}
    \item We present \tool{}, a prototype that automatically identifies AI components in Android apps and generates corresponding AI functionality summarizations.
    \item We curate datasets covering both AI apps and fine-grained AI components to support future research.
    \item We conduct the first analysis to uncover prevalent AI domains and tasks within Android apps.
\end{itemize}

% The prototype implementation and its dataset are publicly available at~\href{https://zenodo.org/15522834/}{https://zenodo.org/records/15522834}.
% We also demonstrate the usefulness of AI service reports generated from \tool{} from the developer's perspective. 

\begin{figure*}[!ht]
    % \james{maybe we need to redraw this figure a bit. in the abstract and introduction, we mentioned that we employed program analysis and LLMs for the four tasks, but we didn't see them in this Figure.}
    \centering
    \includegraphics[width=\linewidth]{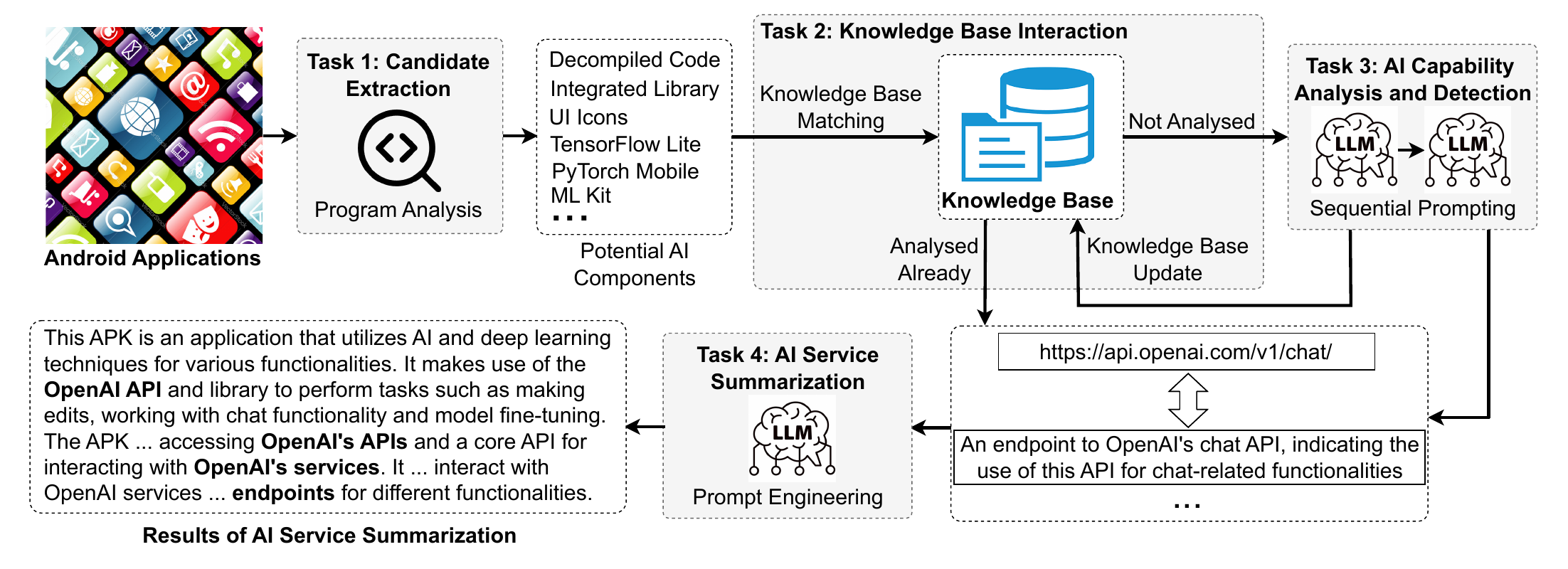} % GPTAIDCP.pdf
    \caption{The workflow of~\tool{}.} %$S_1$: Candidate Extraction, $S_2$ and $S_5$: Knowledge Base Interaction, $S_3$ and $S_4$: AI Capability Analysis and Detection, $S_6$: AI Service Summarization. 
    \label{fig:workingprocess}
\end{figure*}
\section{Approach}
% \sd{I think we can add more static analysis content into the current approach. e.g., how we specifically extract information from apks? }
% \sd{We need to streamline the framework with a clearer fluency. 1. re-draw the overview figure. 2. re-write the context, changing from 6-step to 3/4 successive modules.}
% LLM offers massive neural knowledge to query for all research and development practitioners.
% We integrate static analysis with LLM to identify AI apps, the detailed types of AI, and even further where the identified AI is used. 
% Figure~\ref{fig:workingprocess} shows the whole working process along with some examples adopted in this research, including AI candidate extraction and AI determination. 
% AI candidate extraction is leveraged for necessary candidate information extraction, such as package names, asset names, constant strings, and APIs, via static analysis, consequently, the extracted information is fed into LLM by carefully designed prompts to confirm AI components or services, which is referred to as AI determination. 
% Finally, a summary concludes what type of AI and where it is used will be provided. 

% step-0: AI app filter
% step-1: Candidate Extraction
% step-2: KB Interaction
% step-3: Capability analysis
% step-4: AI candidate detection
% step-5: KB update
% step-6: AI service summarization

Fig.~\ref{fig:workingprocess} illustrates the general framework and workflow of \tool{} (LLM for AI Discovery). 
As mentioned above, the~\tool{} framework consists of four stages. %: candidate extraction, knowledge base interaction, AI capability analysis and detection, and AI service summarization. 
These stages are designed to address the complexity of uncovering AI functionality in Android apps. \textit{Candidate extraction} employs static analysis techniques to extract potential AI components from an Android APK. \textit{Knowledge base interaction} aims to filter these extracted components using a constructed AI and non-AI knowledge base, thereby improving the accuracy of downstream AI capability analysis. The \textit{AI capability analysis and detection} stage further identifies and characterizes AI components by prompting an LLM to analyze. Finally, \textit{AI service summarization} automatically generates a report of these AI components for the given Android APK. Importantly, our approach is model-agnostic and can be applied to any advanced large language model. We describe each stage in the following subsections. 

\begin{figure}[!ht]
    \centering
    \resizebox{0.85\linewidth}{!}{
    \includegraphics[width=\linewidth]{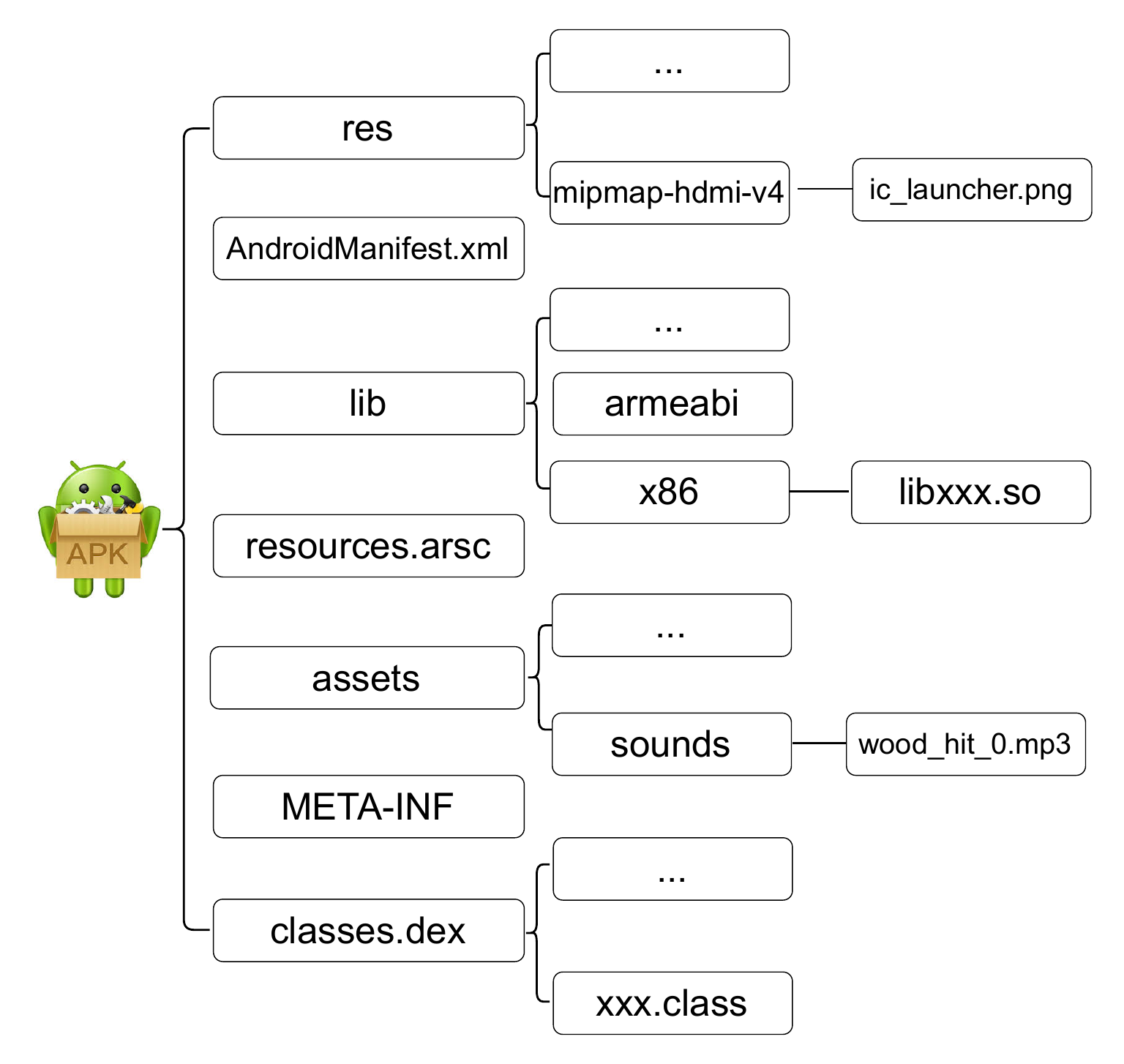}} %android_apk
    \caption{The structure of a typical released Android APK.}
    \label{fig:androidapk}
\end{figure}

% These tasks are achieved by combining static program analysis with the powerful AI advance, ChatGPT, one of the most popular discussed LLMs. 
%To perform these tasks in a chaining style, we utilize ChatGPT with the help of static analysis. 
% Specifically, we designed three offline steps (steps 1-2 and step 5) and three online steps (steps 3-4 and step 6). 
% Importantly, it is worth mentioning that our approach is not limited to ChatGPT and can be applied to any advanced model that has in-context learning capability.

% Fig.~\ref{fig:workingprocess} presents the overall framework and workflow of our approach, i.e., \tool{} (LLM for AI Discovery). Specifically, the AI Discovery process comprises four key tasks: candidate extraction, AI candidate determination, AI capability analysis and detection, and AI service summarization. To accomplish these tasks, we 

\subsection{Task 1: Candidate Extraction}

%After locating the potential apps that use AI techniques, 
To determine AI capabilities within an Android application, we first extract all relevant information—referred to as candidates—from the APK file. Fig.~\ref{fig:androidapk} shows the structure of a typical released Android APK file, typically including classes.dex, which is the Dalvik executable file~\cite{dalvik:2022}, assets, META-INFO, lib, AndroidManifest.xml, res, and resources.arsc. 
Notably, an APK may contain multiple~\texttt{.dex} files due to the method limit of 65,536 per file. 
% When this threshold is exceeded, the compiler automatically splits the code into multiple files, such as classes1.dex, classes2.dex, and so on.
% It is also worth mentioning that the number of the~\texttt{.dex} file in an Android APK may be greater than one. 
% That is because the number of methods contained in the~\texttt{.dex} file is limited to 65536. 
% If the number of methods exceeds 65536 in a~\texttt{.dex}, the Android compiler would split the~\texttt{.dex} file into two, and name them classes1.dex and classes2.dex.
% To identify the structure of the Android APK under analysis, we utilize AAPT~\cite{aapt:2024}, an official Android build tool widely used in Android Studio and the Gradle plugin. 
We utilize AAPT~\cite{aapt:2024} to extract the structure from the curated Android APK file.
Our focus is on examining the implementation details embedded in the app, which requires decompiling \texttt{.dex} files to retrieve the underlying Java logic and analyzing \texttt{.so} files in the lib directory for native code.
% In this research, we are interested in the details of the implementation of the APK. 
% We are supposed to decompile the~\texttt{.dex} file to visit the underlying Java implementation details and also the dynamic libraries ending with the suffix~\texttt{.so} to inspect the native code implementations in the~\texttt{lib} directory. 
Specifically, we extend a static analysis approach, namely App Scanner, proposed by Liu~\etal{}~\cite{liu2021identifying} to extract APIs from an APK file. The scanner aims to transform Dalvik bytecode~\cite{dalvik:2022} into Jimple~\cite{bartel2012dexpler}, a human-readable intermediate representation. 
% was implemented based on the popular static analysis framework, Soot~\cite{lam2011soot}, which could convert the Dalvik bytecode~\cite{dalvik:2022} of APK into Jimple~\cite{bartel2012dexpler} (the default internal representation in Soot and readable for humans).  
%The traditional Java-based static analysis framework, Soot, was enhanced with the ability to convert Dalvik bytecode into Jimple's three-address code~\cite{bartel2012dexpler}.  Dalvik bytecode~\cite{dalvik:2022} is the default executable representation for Android apps. It is indirectly compiled from Java code due to the running environment with less memory and power provision, while Jimple is the default internal representation in Soot. 
The App Scanner method is enhanced to additionally capture fine-grained components from multiple sources within the given APK, such as \texttt{.dex} files, native library~\texttt{.so} files, and the asset files. This also allows us to extract more indicators of AI functionality, such as HTTP requests~\texttt{https://api.openai.com/v1/} and descriptive package name~\texttt{com.google.firebase.ml.vision.o-\\bjects}.
Moreover, to analyze embedded dynamic native libraries, we employ tools \texttt{readelf}~\cite{readelf:2023} and \texttt{objdump}~\cite{objdump:2023} to parse HTTPS requests, as many apps implement AI capabilities by invoking these requests via the Android NDK~\cite{ndk:2023}.

% Concerning the integrated dynamic native libraries, we also try to parse HTTPS requests from these with the help of~\texttt{readelf}~\cite{readelf:2023} and~\texttt{objdump}~\cite{objdump:2023} as some apps implement AI capabilities in native code and invoke such inferences via NDK~\cite{ndk:2023}.
\begin{figure}[!ht]
    \centering
    \includegraphics[width=0.98\linewidth]{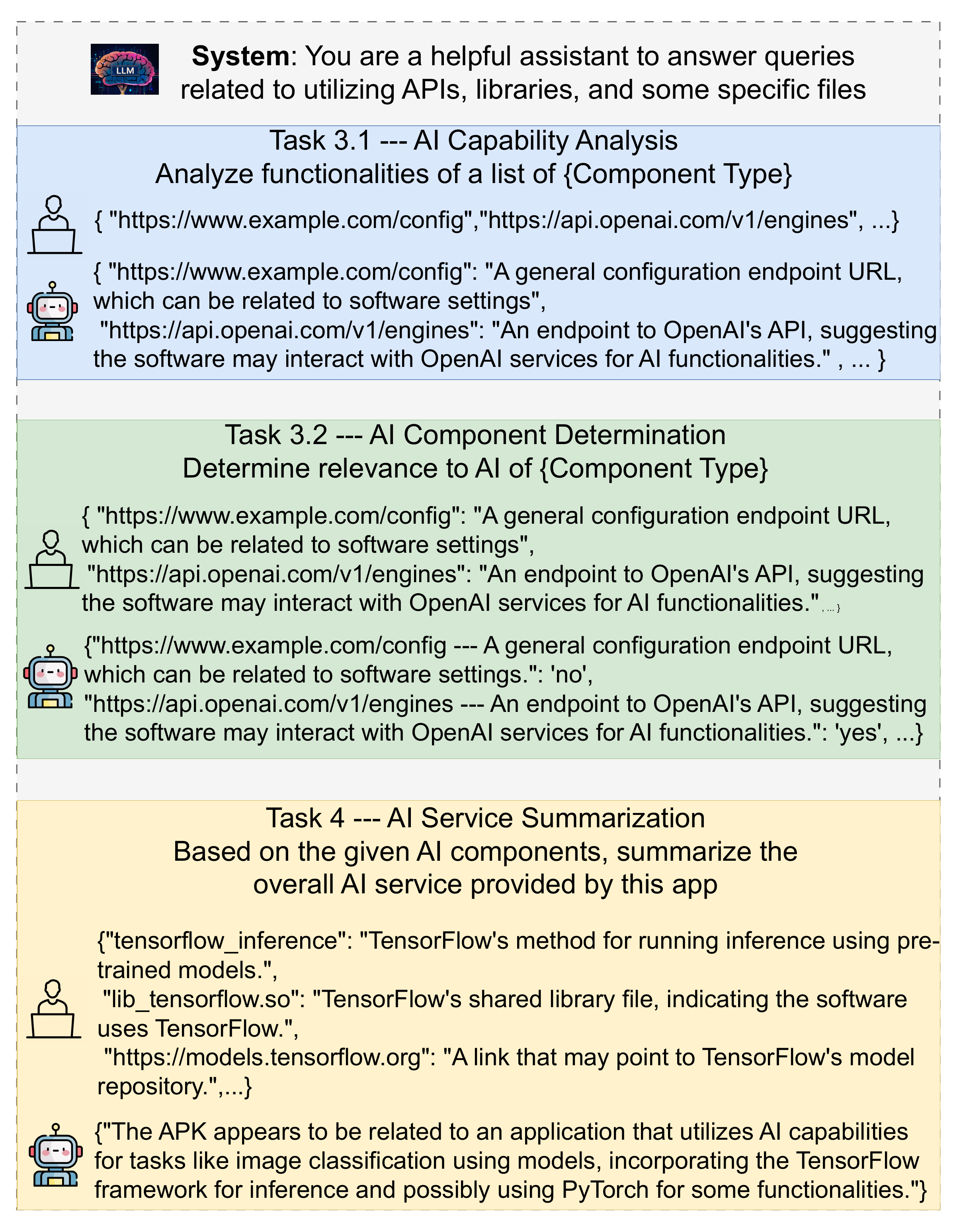}
    \caption{Prompt design and examples for \tool{}}
    \label{fig:enhanced_prompt}
\end{figure}

\subsection{Task 2: Knowledge Base Interaction}
% \jiawei{I reckon we need a more clear description of how LLMs interact with our AI and non-AI Knowledge base in this section. From my understanding we used the in-context learning strategy with few-shot examples to do this. Our AIKB is composed of two-parts: AI KB and non-AI KB, these are all json files. Our current approach is based on in-context learning with few-shot examples, I wonder if it would be better if we can leverage more SOTA approaches such as RAG to do this query? This seems to be more dynamic and extensible than few-shot in-context learning? RAG+LLaMA3 (comparison)}
% Given our focus on the implementation-level details of AI capabilities, the preceding extraction process produces a substantial number of fine-grained components that capture relevant aspects of these implementation details. 
The extraction process yields a substantial number of fine-grained components capturing both AI and non-AI implementation details.
To filter out the irrelevant non-AI components, we apply a rule-based approach following the extraction process. 
This approach aims to remove components that are irrelevant to AI functionalities; however, it operates at a coarse granularity, which may result in the retention of some non-AI components falsely classified as AI-related.
% This approach is designed to eliminate components that do not exhibit characteristics indicative of AI-related functionality, thereby reducing noise and improving the precision of subsequent analysis. 
% However, the rule-based method we designed operates at a coarse granularity, which means that some components initially classified as AI-related may be irrelevant.
To further refine the analysis and enhance efficiency, we incorporate a knowledge base (KB) interaction step into the workflow. This step enables the reuse of prior analysis results to accelerate processing and reduce redundant computation. Specifically, we maintain two separate KBs: an AI KB that stores previously verified AI components and their associated capability analyses, and a non-AI KB that contains confirmed non-AI components used primarily for pre-analysis filtering. Once the components are extracted and initially filtered, they are checked against both KBs. If a match is found, the corresponding analysis and detection result are retrieved directly, bypassing further computation. 
Otherwise, the components are forwarded to the next stage, Task 3: AI capability analysis and detection, whose results are then systematically incorporated into the corresponding KB for future reference and reuse.
%The components' results of this stage are systematically incorporated into the corresponding KB for future reference and reuse.
% This requires subsequent analysis steps to further verify their relevance. 
% To accelerate future analyses when encountering the same components and reduce computational cost, we introduce a knowledge base interaction step into the workflow.
% We build a knowledge base (KB) to store all components that have already been analyzed in prior processed apps, so as to speed up the determination of newly extracted components. 
% Specifically, we build two KBs, AI KB (for genuine AI components) and non-AI KB (for authentic non-AI components), to store all components and their analysis results. 
% While AI KB aims to help provide all capability analysis for AI service summarization, non-AI KB mainly helps reduce the number of irrelevant components before analysis. 
% Once the implementation details are extracted from the given APK file and immediately filtered with the rule-based approach, they will be examined in both KBs afterwards. 
% The corresponding analysis and detection results will be returned directly if they have been analyzed. 
% Otherwise, they will be passed on to the next task (\ie{}, Task 3: AI capability analysis and detection), and then these components along with their analysis results will be systematically incorporated into the corresponding knowledge base for future reference and reuse.

\subsection{Task 3: AI Capability Analysis and Detection}
This stage, the core of \tool{}, involves identifying and characterizing the extracted AI components. To ensure precision and interpretability, this task is divided into two sub-steps: \textit{AI capability analysis} and \textit{AI component determination}. After passing through the knowledge base interaction phase, we obtain a set of deduplicated candidate components that are used for further analysis. For each component, we issue an initial prompt to ChatGPT to produce a functional-oriented interpretation.
% Regarding each component, an initial prompt is issued to ChatGPT to elicit a functional-oriented analysis. 
For example, given the package name~\texttt{com.google.mlkit.vision.objects}, ChatGPT may provide the analysis result of ``Google's ML Kit API for object detection and tracking in images and videos.'' 
In cases where components are obfuscated or highly customized, \ie{}, \texttt{nlp.WordPieceModelPB}, ChatGPT can still offer meaningful functional descriptions, \eg{}, ``A library for tokenizing text using the WordPiece algorithm, implemented in Protocol Buffers format'' which would be difficult to identify using a rule-based method approach. 
% insights based on the contextual patterns and the given description. 
% For instance, the package \texttt{nlp.WordPieceModelPB} can be hard to determine with rule-based methods. 
% With the help of ChatGPT, the human-like speculation of `A library for tokenizing text using the WordPiece algorithm, implemented in Protocol Buffers format' can effectively describe the highly potential functionality. 
Based on the generated analysis, we further prompt ChatGPT to determine whether the component falls within the scope of AI and justify its decision with a detailed explanation. 
Leveraging ChatGPT's ability to interpret and reason over its own generated content, the sub-steps of AI capability analysis and AI component determination can be straightforwardly and effectively conducted by adopting the prompts of Task 3.1 and Task 3.2 in Fig.~\ref{fig:enhanced_prompt}. 
% \subsection{Step 4: AI Candidate Detection}

\subsection{Task 4: AI Service Summarization}
% \sd{I think instead of some textual description as summarization, we can provide more structure information. Maybe something like privacy labels? The content displayed in table-like format.}
Given the determined AI components and their corresponding functional descriptions, we generate a concise AI service report by employing a a well-crafted ChatGPT prompt to summarize the AI capabilities of these components. 
% a brief report describing what AI services the mobile apps offer could be generated via a well-crafted ChatGPT prompt. 
% However, since ChatGPT and most LLMs always have limited context length, it is challenging to put all the determined AI component names and their corresponding analysis into the same context window.
However, due to the inherent context length limitations of ChatGPT and  LLMs in general, incorporating all component names and their analyses into a single prompt is often infeasible.
% To address this limitation, we exclude the API-level components, as their functionalities are likely to be implied by their packages. 
To mitigate this constraint, we omit low-level API components, \ie{}, their functionalities are typically inferred from their package context, and focus on higher-level components instead.
By concatenating the remaining AI-relevant components, we prompt ChatGPT to generate the AI service report for app users, which is shown in Task 4 of Fig.~\ref{fig:enhanced_prompt}. 
% It is also worth mentioning that the report can be tailored to different audiences, such as app users, app regulators, and app developers, by slightly adapting the prompt, thereby enabling context-specific interpretations and supporting diverse analysis needs. 
Notably, the generated report can be adapted for different audiences, such as end users, regulators, or developers, by slightly adjusting the prompt, thereby enabling context-specific interpretations and supporting diverse analysis needs.

\subsection{Experiment Configuration}
To instruct ChatGPT to execute each sub-step precisely, we carefully design natural language prompts, which are presented in Fig.~\ref{fig:enhanced_prompt}. To make ChatGPT fully understand these analysis steps, we additionally provide five input-output examples for the few-shot in-context learning~\cite{brown2020language,wang2020generalizing}. 
In our implementation of~\tool{}, we utilize the default 4K-token context window provided by the ChatGPT API. Given  ChatGPT's rate-limiting, processing one app with one component per request becomes time-consuming and inefficient. Moreover, submitting too many components in a single request may lead to output truncation or omission, resulting in misalignment between input and output. 
% However, including too many components in single request may cause ChatGPT to omit certain items in the output, resulting misalignment. 
To balance task accuracy and processing efficiency, we adopt a batch size of three components per request and instruct ChatGPT to return structured outputs in JSON format. Model parameters are left at default values, except for the temperature and TopP settings, which are adjusted from 1.0 to 0.2 and 0.95, respectively. These adjustments reduce output variability while retaining a certain degree of creativity, thereby improving the consistency and reliability of the component analyses and summary reports.
% The model parameters are primarily kept at their default settings, with the exception of the temperature and TopP values, which are adjusted from 1.0 to 0.2 and 0.95, respectively. 
% This adjustment is intended to reduce output randomness while preserving a degree of creativity, thereby supporting the generation of accurate analysis results and meaningful app summaries.
\section{Evaluation}
% \sd{Maybe add another RQ which is directly related to AI security? \eg{}, lets say someone points out there is a xxx problem about GPT, and we can estimate how serious of this problem in practice.}
In this section, we evaluate the effectiveness of our approach, \tool{}, by addressing the following research questions:
% \sd{We can combine the RQ1, RQ2, and RQ3. The new RQ2 would be the pattern of accessing to the AI in mobile apps. The new RQ3 would be the estimation of severity of security and privacy risks in mobile apps.}
% \jiawei{Google play apk metadata is sometimes insufficient to get what AI features are used in collected apps. User reviews are more straightforward for trend/pattern analyses of AI usage in modern android apps. Existing AI features and AI features people request can provide insights into future work and potentially serve as a case study of LLMAID analysis results and real-world trend.}

\begin{itemize}[listparindent=\parindent]
    \item \textbf{RQ1:} To what extent can \tool{} reliably and accurately determine AI capabilities in previously identified AI Apps?\\ % How reliable is \tool{} in determining AI capability for existing AI apps?
    \indent This research question aims to assess the accuracy and reliability of \tool{} in analyzing AI capabilities within previously recognized AI Android apps. 
    % Additionally, we explore the possibility of making \tool{} more efficient and discuss the trade-offs between processing time and result reliability. 
    % By answering this research question, we confirm the efficacy of~\tool{} and provide the research basis for the following study. 
    Addressing this question validates the core effectiveness of \tool{} and establishes a foundation for further evaluation.
    
    \item \textbf{RQ2:} How does \tool{} compare to the state-of-the-art rule-based approach in reliably identifying AI capabilities within in-wild Android Apps? \\ % How reliable is \tool{} in identifying AI capability within large-scale wild Android apps?
    \indent We compare the performance of~\tool{} with that of the SOTA rule-based method, i.e., \textit{AI Discriminator}, in identifying AI capabilities across large-scale Android apps. By answering this research question, we demonstrate the effectiveness of~\tool{} in accurately determining AI capabilities.
    %, thereby enhancing the detection of AI-enabled Apps. 

    \item \textbf{RQ3:} How is the usefulness of the AI service summary generated by~\tool{}? \\
    \indent We explore the clarity and informativeness of the AI service summary provided by our framework. This summary provides end users, developers, and regulators with an insightful overview of the AI functionalities of a given Android app.
    % In addition to pinpointing AI components of Android apps, we also sum up the possible AI capabilities regarding the determined AI components in this research question. By answering this research question, we could gain a clearer understanding of what types of AI technology have been incorporated. 
    %and provide a more meaningful and insightful summary for future app developers. %, and even for future regulators and any interested stakeholders. 

    \item \textbf{RQ4:} To what extent can \tool{} uncover prevalent AI capabilities and usage patterns in real-world Android Apps? \\ % What insights can be gained from the \tool{} analysis?
    \indent  This final research question investigates the broader landscape of AI adoption in mobile apps by analyzing the types and distributions of AI tasks and domains identified by \tool{}. The findings offer a deeper understanding of the AI functionalities embedded in these apps and provide valuable insights to guide AI researchers and developers in designing and improving their app features.
    % \indent In this last research question, we further categorize the AI-related tasks and functional domains associated with the identified AI components in the determined AI Apps. By answering this research question, we could gain a deeper understanding of the specific AI functionality provisions in AI apps, offering development suggestions for downstream App developers and valuable insights for AI researchers.
\end{itemize}

\subsection{Dataset}
We perform our experiments on two separate datasets. The first one comes from the existing literature, which was collected from AndroZoo and contains 56,682 rule-based approach determined AI Android apps~\cite{allix2016androzoo}. The other one is self-curated, comprising 15,517 Android apps, to address the absence of recently released applications, especially those integrated with emerging LLMs such as ChatGPT. We present these two datasets in the following paragraphs. 

\subsubsection{Dataset 1}
Li~\etal{}~\cite{li2022ai} proposed a SOTA rule-based approach, \textit{AI Discriminator}, to identify AI-driven Android apps via keywords matching on the dissected and decompiled Android apps. They successfully identified 56,682 AI apps from a total of 7,259,232\footnote{The latest Android version is considered while multiple versions are stored.} apps on AndroZoo. In our experiments, we utilize the dataset of 56,682 Android AI apps to perform cross-validation of our approach.

% \begin{figure} [t!]
%     \centering
%     \includegraphics[width=0.96\linewidth]{res/app_dist.pdf}
%     \caption{The distribution of collected apps over the years.}
%     \label{fig:appdist}
% \end{figure}

\subsubsection{Dataset 2}
To curate our dataset, we leverage AndroZoo and extract the latest release of each Android app using metadata from its associated CSV file. For each app, we utilize the google-play-scraper npm package to verify its current availability on Google Play Store by querying with the package name~\cite{playscraper:2023}, 
%If the app is still publicly accessible, we
and collect its key metadata, particularly the app description and release date.
% To curate our dataset, we utilize AndroZoo and extract the latest release for every Android app from its provided CSV file.
% With these latest app releases, we resort to the npm package google-play-scraper~\cite{playscraper:2023} by feeding the package name to it to determine if the app is still publicly available on Google Play store and, if so, to retrieve the app metadata, especially the app description and release date. 
In total, we identified 2,298,423 actively maintained Android apps out of a total of 23,321,823 apps available in AndroZoo at the time of our study. To focus our evaluation on apps with likely AI functionality, we further refined the dataset by examining the app descriptions retrieved from Google Play, selecting only those that strongly indicated the presence of AI-related features.
% In total, we identified 2,298,423 actively maintained Android apps among a huge number of 23,321,823 apps on AndroZoo (as of the time of our study), and also we further filtered the highly possible potential AI apps (\ie{}, those are very likely to be AI Apps) from them by checking their description crawled from Google Play Store to tailor the number of experimental apps.
%as we focus on . % the reason why do we need to filter out non-AI apps

Previous studies often rely on manual inspection to identify AI apps, however, this approach tends to be time-consuming and challenging to scale. 
%. However, this process is time-consuming. 
% To fasten the possible AI app identification, we explore the possibility of determining whether an app uses AI techniques or not without detailed inspection. 
A publicly available Android app conventionally includes a brief description that clarifies its core functionalities. For example, the description of Wise AI Chatbot\footnote{\url{https://play.google.com/store/apps/details?id=com.jgapps.chatgpt}} mentions that `\textit{Wise Chat GPT Pro is the first app to utilize ... ChatGPT 3.5 Turbo to provide instant and \textcolor{red}{[intelligent]} responses ... with Wise Chat GPT Pro, our \textcolor{red}{[AI] [chatbot]} powered by ChatGPT 3.5 Turbo technology ...}'. This example illustrates how AI-related apps often explicitly reference AI-related terms (i.e., highlighted in red) in their descriptions. 
% Such AI-related apps usually use AI-related terms (\ie{}, highlighted ones) in the description. 
Motivated by this observation, we transform this step into a two-staged process. We begin by applying a rule-based filter to identify potential AI apps through the presence of AI-related keywords in their descriptions. Subsequently, we employ ChatGPT to perform a deeper semantic analysis of these filtered descriptions, further verifying whether the apps likely incorporate AI technologies. %We present the two-stage process in the following paragraph. 
% , where we sequentially perform the rule-based filtering and apply ChatGPT to identify the apps potentially leveraging AI technologies.

To be specific, the filtering stage employs rule-based keyword extraction to identify potential AI apps based on their application descriptions. We define a wide range of keywords from high-level functionalities (\eg{}, ``AI'' and ``Deep Learning'') to low-level ones (\eg{}, ``text classification'' and ``facial recognition''). By matching these terms exactly, the filter effectively reduces the candidate pool, allowing subsequent analysis to focus on high-potential AI apps. However, the mere presence of such keywords does not always indicate that AI constitutes a core functionality of the app. To address this limitation, we introduce a second filtering stage using ChatGPT, which evaluates whether the app is likely to employ AI techniques based on its description. This stage further enhances the precision of our identification process by interpreting the context of the apps' descriptions. By employing the two-stage process, our approach overcomes the limitation of keyword matching methods, hence accurately detecting AI Android apps. 

% However, the presence of predefined terms in descriptions does not necessarily indicate that they reflect the app's primary functionalities. Thus, we incorporate ChatGPT as a second filter to classify if the app will likely use AI techniques based on the given description. % by following the sample prompts in Fig.~\ref{fig:enhanced_prompt}. 
% %The corresponding prompt is shown in Step 0 in Figure~\ref{fig:enhanced_prompt}. 
% The integration of ChatGPT addresses the limitations of rule-based methods by adding a layer of semantic depth, which is essential for the comprehensive identification of AI-integrated applications.

\begin{figure*}[!ht]
    \centering
    \begin{minipage}[t]{0.48\linewidth}
        \centering
        \includegraphics[width=\textwidth]{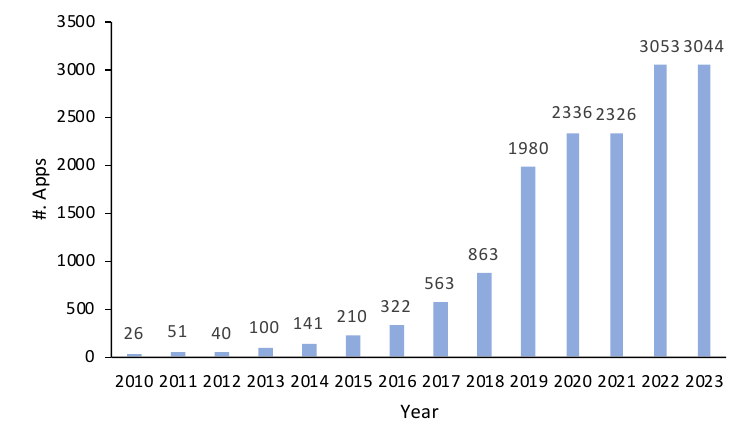}
        \caption{The distribution of collected apps over the years.}
        \label{fig:appdist}
    \end{minipage}
    \begin{minipage}[t]{0.48\linewidth}
        \begin{minipage}[t]{0.46\textwidth}
            \centering
            \includegraphics[width=\textwidth]{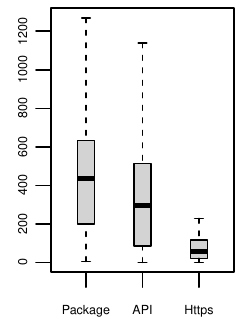}
            \subcaption{Considered.}
            \label{fig:total_pkg_api_str}
        \end{minipage}
        \begin{minipage}[t]{0.46\textwidth}
            \centering
            \includegraphics[width=\textwidth]{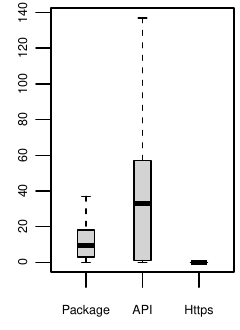}
            \subcaption{Determined.}
            \label{fig:deter_ai}
        \end{minipage}
        \caption{The determination of AI candidates.}
        \label{fig:rq1_pkg_api_str}
    \end{minipage}
\end{figure*}

In total, we successfully identified 15,517 Android apps (only the latest released version is considered) from a total of 23,321,823 apps on AndroZoo. 
Additionally, we extracted the release date from the metadata provided on the Google Play Store and presented the distribution of these apps over the last decade in Fig.~\ref{fig:appdist}. Our data shows that the growth in released apps aligns closely with the increasing prevalence of AI technologies during this period.
% We observe that the rise of the released apps is at the same pace as the prevailing AI technologies. 
Notably, Fig.~\ref{fig:appdist} includes only 15,055 apps, as release date metadata was unavailable for 462 apps, representing just 2.98\% (462 out of 15,517) of the total. Despite this, the upward trend of employing AI technologies in mobile apps over the past ten years remains valid. 
% we find that the total number of apps in Fig.~\ref{fig:appdist} is 15,055 due to the release date missing in the given metadata among 462 Android apps, which only accounts for 2.98\% (462/15,517).
% Therefore, the rising trend of AI apps over the past ten years is still valid.

\subsection{\textbf{RQ1:} To what extent can \tool{} reliably and accurately determine AI capabilities in previously identified AI Apps?}

% \begin{figure}[t!]
%     \centering
%     \begin{minipage}[t]{0.23\textwidth}
%         \centering
%         \includegraphics[width=\textwidth]{res/pkg_apk_str_total.pdf}
%         \subcaption{Considered.}
%         \label{fig:total_pkg_api_str}
%     \end{minipage}
%         \begin{minipage}[t]{0.23\textwidth}
%         \centering
%         \includegraphics[width=\textwidth]{res/pkg_apk_str_ai.pdf}
%         \subcaption{Determined.}
%         \label{fig:deter_ai}
%     \end{minipage}
%     \caption{The determination of AI candidates.}%, including but not limited to packages, APIs, and strings.}
%     \label{fig:rq1_pkg_api_str}
% \end{figure}

In this research question, we examine whether \tool{} can effectively identify AI capabilities within a set of pre-labeled AI applications. 
%In this research question, we intend to figure out if the different orders of reasoning and classification would have different impacts on the final AI determination results and hence have more accurate analysis results. 
%If so, we would like to select the order producing higher accuracy in AI App determination and construct it into our final LLM prompt. 
%For evaluation, we utilize \textit{Dataset1}. % which contains 56,682 AI apps curated by Li~\etal{}~\cite{li2022ai}. 
We employ \textit{Dataset 1} as our evaluation benchmark. 
However, since \textit{Dataset 1} only determines AI apps but does not provide ground-truth labels for their detailed underlying AI components, we randomly sample a subset of 382 AI apps and perform human validation to assess \tool{}. 
The sample size is calculated based on the popular online Sample Size Calculator~\footnote{\url{https://www.surveysystem.com/sscalc.htm}} with a confidence level of 95\% and a margin of error of 5\%. 

We first present the quantity distribution of three representative types of components, packages, APIs, and HTTPS requests in Fig.~\ref{fig:rq1_pkg_api_str}. 
Specifically, Fig.~\ref{fig:total_pkg_api_str} shows the total number of packages, APIs, and HTTPS requests after completing \textit{Task 1} (Candidate Extraction) of \tool{}, while
Fig.~\ref{fig:deter_ai} presents the identified AI-related packages, APIs, and HTTPS requests after finishing \textit{Task 3} (AI capability analysis and detection) of \tool{}. 
% We present other statistics in \cite{anonyzenodo}.
The median values for the total considered packages, APIs, and HTTPS requests in Fig.~\ref{fig:total_pkg_api_str} are 433.5, 296, and 53.5, respectively. 
It is also worth noting that the considered packages, APIs, and HTTPS requests are proactively filtered results obtained by applying a self-curated whitelist of commonly observed non-AI components during the candidate extraction process (\ie{}, Task 1). 
The whitelist of non-AI components comprises the official Java and Android kernel-related packages and API signatures. 
%The remaining ones shown in the boxplots are those that are likely to be AI-related and we cannot filter them out by rule-based filters. 
% After the process of AI capability analysis and detection, we can get the final number of AI-related packages, APIs, and strings in Figure~\ref{fig:deter_ai}. 
After performing AI capability analysis, the median number of AI-related packages, APIs, and HTTPS requests drops to 9.5, 33, and 0, respectively. Notably, packages are filtered more aggressively than APIs, which is expected given that many AI-specific APIs are concentrated within a smaller set of AI-related packages. In contrast, HTTPS requests are sparsely detected, primarily due to the relatively recent adoption of AI-related network calls and their limited presence in the collected dataset.
% The median values for the identified AI packages, APIs, and HTTP requests are 9.5, 33, and 0, respectively. 
% We observe that more packages are filtered than APIs, which can be attributed to the fact that many AI-related APIs are concentrated within a limited number of AI-specific packages.
% In terms of HTTPS requests, they are only detected sparsely by~\tool{}. 
% That is because the introduction of HTTP requests for OpenAI is comparatively late and they are not pervasive among the collected experimental apps. 

To evaluate the reliability of \tool{} in identifying AI components within the sampled dataset, we conducted a manual annotation study involving two independent authors, who independently labeled a total of 49,371 components extracted from the 382 sampled AI apps. When disagreements arise, dedicated meetings involving an additional author are held to facilitate consensus. Each component was classified as either an AI component or a non-AI component. We defined a \textit{True Positive} as a component correctly identified by \tool{} as AI-related, while a \textit{True Negative} referred to a component correctly identified as non-AI. Following human validation, \tool{} achieved a precision of 98.05\% and a recall of 93.31\%, indicating strong performance in correctly detecting AI components. Furthermore, inter-annotator agreement measured using Cohen’s unweighted kappa coefficient~\cite{cohenkappa} was 0.82, reflecting substantial consistency among human evaluators. These results collectively demonstrate the high accuracy and robustness of \tool{} in detecting AI components.

% To examine the reliability of \tool{} in determining AI components in the sampled dataset, we asked multiple authors to independently label 49,371 components extracted from the sampled 382 AI apps. 
% Components are classified into two groups: AI components and non-AI components. We define True Positive as the components that are correctly identified as AI components, while True Negative refers to the components that are correctly detected as non-AI components. 
% After human validation, we achieved a high precision of 98.05\% and recall of 93.31\%, with Cohen's unweighted kappa coefficient~\cite{cohenkappa} of 0.82. The results indicate that \tool{} is highly accurate in identifying AI components. % and external observers are very likely to agree on the \tool{} outcomes.

In addition, we validated the reliability of \tool{} in identifying AI apps. We defined an app is considered an AI app if it contains one or more AI components determined by \tool{}. 
Our results showed that \tool{} can detect 378 out of 382 sample apps as AI apps, indicating an accuracy of 99.0\% (372/382) is achieved. We further analyzed the remaining four apps and observed that the only potential AI indicators were filenames with the \texttt{.model} suffix. 
Without further complementary information, such as associated APIs, packages, or functional descriptions, \tool{} was unable to classify these apps as AI apps, hence excluding them from the set of AI apps. 

\begin{table}[]
    \centering
    \caption{Comparison of the reliability of~\tool{} between the implementations of \textit{analysis-then-detection} and \textit{detection-then-analysis}.}
    \begin{threeparttable}
    % \resizebox{0.96\linewidth}{!}{
    \begin{tabular}{c|c|c|c|c|c}
    \hline
    \multicolumn{3}{c|}{Analysis/Detection} &  \multicolumn{3}{c}{Detection/Analysis} \\
    \hline
       \makecell{AI Candidate}  &  \makecell{AI Apps} & \makecell{Time (sec)\tnote{1}} & \makecell{AI Candidate}  &  \makecell{AI Apps} & \makecell{Time (sec)}\\
    \hline
          2859  & 378       & 638.76  & 2192 &    371   & 226.94    \\
    \hline
    \end{tabular}
    % }
    \begin{tablenotes}
    \footnotesize
    \item[1] Average running time for an Android APK in seconds.
    \end{tablenotes}
    \end{threeparttable}
    \label{tab:rq1_reason_class}
\end{table}
% In the evaluation process of \tool{}, we find that \tool{} is relatively slow since it first needs to generate the analysis for both AI and non-AI components. 
% Although the design of \textit{Task 3.1} (AI Capability Analysis) and \textit{Task 3.2} (AI Component Determination) is motivated by \textit{Chain-of-Thought}~\cite{wei2022chain}, a prompting technique to elicit reasoning ability from LLMs, we argue that it may not be optimized for \tool{}. 
% To efficiently determine AI capability, an intuitive approach is to first classify components and then provide analysis descriptions only for AI components (\ie{}, exchange the execution order of Task 3.1 and Task 3.2). 
% To evaluate the reliability of this approach, we follow the same evaluation settings among the sampled 382 apps. 
% As shown in Table~\ref{tab:rq1_reason_class}, the processing time per Android app is significantly decreased by 65\% after adopting the \textit{detection-then-analysis} implementation. 
% During the evaluation of \tool{}, we observe that its processing is relatively slow due to the need to generate analyses for both AI and non-AI components. 
Although the design of \textit{Task 3.1} (AI Capability Analysis) and \textit{Task 3.2} (AI Component Determination) is inspired by Chain-of-Thought prompt~\cite{wei2022chain}, it may not be optimal for \tool{}. 
A more efficient alternative is to first classify the components and then analyze only those identified as AI components, reversing the order of Tasks 3.1 and 3.2. 
Following the same evaluation setup on the 382 sampled apps, we find that this detection-then-analysis approach reduces the processing time per app by 65\%, as shown in Table~\ref{tab:rq1_reason_class}.
The number of identified AI components is significantly lower than that of the original sequential implementation. 
% Although the new alternative approach is still capable of detecting most AI apps (371 vs. 378), the number of identified AI components is significantly lower than that of the original sequential implementation. 
% The observation indicates the trade-off between the reliability and efficiency of \tool{}. 
Based on the overall results, we suggest that \textit{detection-then-analysis} can only be used when developers are interested in finding AI apps. If developers seek to understand all possible AI capabilities of an app, we foresee that the original \tool{} with \textit{analysis-then-analysis} is more effective in capturing AI candidates.

\begin{tcolorbox}[before skip=0.4cm, after skip=0.6cm, title=\textbf{RQ1 Findings}, left=2pt, right=2pt,top=2pt,bottom=2pt]
%% add high accuracy of reasoning output and then we have a higher ai app detection outcome.
% We demonstrate the efficacy of \tool{} in determining AI capabilities for existing AI apps. 
Human validation suggests that \tool{} achieves high precision (98.05\%) and recall (93.31\%) in detecting AI components and AI apps. 
%Meanwhile, we also attempt to improve the efficiency of \tool{}. 
Although reversing the sequence of Task 3.1 and Task 3.2 could improve the efficiency of \tool{}, it compromises reliability compared to the original sequential implementation.
\end{tcolorbox}

% \begin{figure}
%     \centering
%     \includegraphics[width=0.5\linewidth]{res/venn_discriminator_discovery.pdf}
%     \caption{Venn diagram of detected AI apps in AI Discriminator and~\tool{}.}
%     \label{fig:venn_diagram}
% \end{figure}

% \begin{table*}
% \begin{minipage}[t]{0.98\columnwidth}
%     \centering
%     \captionof{table}{Number of AI and Non-AI apps classified by AI Discriminator and~\tool{}.}
%     \begin{threeparttable}
%     % \resizebox{0.90\linewidth}{!}{
%     \begin{tabular}{c|c|c|c}
%     \hline
%     \multicolumn{2}{c|}{AI Discriminator} &  \multicolumn{2}{c}{\tool{}} \\
%     \hline
%         AI Apps &  Non-AI Apps &  AI Apps &  Non-AI Apps \\
%     \hline
%            603  &       3216      &    1366    &     2453   \\
%     \hline
%     \end{tabular}
%     % }
%     \end{threeparttable}
%     \label{tab:rq2_deter_aiarison}
% \end{minipage}
% \vspace{6mm}
% \begin{minipage}[t]{0.98\columnwidth}
%     \centering
%     \captionof{table}{Detected AI Apps by AI Discriminator and~\tool{}}
%     % \resizebox{0.90\linewidth}{!}{
%     \begin{tabular}{c|c|c}
%     \hline
%         AI Discriminator &  \makecell{AI Discriminator\\ \tool{}} & \tool{} \\
%     \hline
%          9               &    546                    & 820 \\
%     \hline
%     \end{tabular}
%     \label{tab:venn}
%     % }
% \end{minipage}
% \end{table*}

\subsection{\textbf{RQ2:} How does \tool{} compare to a state-of-the-art rule-based approach in reliably identifying AI capabilities within in-wild Android Apps? }

% In this research question, we aim to evaluate the generalizability of~\tool{} in comparison to the state-of-the-art rule-based approach, AI Discriminator~\cite{li2022ai,aiapp:2023}. 
% Specifically, 
In this research question, we evaluate how reliable \tool{} is in determining AI capabilities and compare it to the SOTA approach, AI Discriminator~\cite{li2022ai,aiapp:2023}. 
% We reuse the implementation of AI Discriminator~\cite{li2022ai} from~\cite{aiapp:2023}. 
To efficiently perform the evaluation, we select a representative subset of \textit{Dataset 2}, focusing on the apps published in the recent five years, and randomly sampling 30\% apps in each year to preserve the same quantity distribution.
Finally, we ended up with 3,819 $(1980 \times 30\% + 2336 \times 30\% + 2326 \times 30\% + 3053 \times 30\% + 3044 \times 30\%)$ apps for the experiment.

Following the same experimental setup as in RQ1, 
we compute the precision and recall for AI component identification, and 
\tool{} achieves a precision of 97.32\% and a recall of 91.01\% in identifying AI components.  
The inter-annotator reliability, measured by Cohen's unweighted kappa coefficient, is 0.80, indicating a high level of agreement with the results generated by~\tool{}. 

% \begin{table}[]
%     \centering
%     \caption{Number of AI and Non-AI apps classified by AI Discriminator and~\tool{}.}
%     \begin{threeparttable}
%     \resizebox{0.98\linewidth}{!}{
%     \begin{tabular}{c|c|c|c}
%     \hline
%     \multicolumn{2}{c|}{AI Discriminator} &  \multicolumn{2}{c}{\tool{}} \\
%     \hline
%         AI Apps &  Non-AI Apps  &  AI Apps  &  Non-AI Apps \\
%     \hline
%            603  &       3216      &    1366    &     2453   \\
%     \hline
%     \end{tabular}}
%     \end{threeparttable}
%     \label{tab:rq2_deter_aiarison}
% \end{table}

% \begin{table}[]
%     \centering
%     \caption{Detected AI Apps by AI Discriminator and~\tool{}}
%     \resizebox{0.98\linewidth}{!}{
%     \begin{tabular}{c|c|c}
%     \hline
%         AI Discriminator &  \makecell{AI Discriminator\\ \tool{}} & \tool{} \\
%     \hline
%          9               &    546                    & 820 \\
%     \hline
%     \end{tabular}
%     \label{tab:venn}}
% \end{table}

% Like RQ1, we also provide an evaluation of AI app determination. 
% Table~\ref{tab:rq2_deter_aiarison} presents the detection results in terms of the rule-based approach AI Discriminator and our tool~\tool{}. 

\begin{table}[!ht]
    \centering
    \captionof{table}{Number of AI and Non-AI apps classified by AI Discriminator and~\tool{}.}
    \begin{threeparttable}
    % \resizebox{0.90\linewidth}{!}{
    \begin{tabular}{c|c|c|c}
    \hline
    \multicolumn{2}{c|}{AI Discriminator} &  \multicolumn{2}{c}{\tool{}} \\
    \hline
        AI Apps &  Non-AI Apps &  AI Apps &  Non-AI Apps \\
    \hline
           603  &       3216      &    1366    &     2453   \\
    \hline
    \end{tabular}
    % }
    \end{threeparttable}
    \label{tab:rq2_deter_aiarison}
\end{table}

\begin{table}[!ht]
    \centering
    \captionof{table}{Detected AI Apps by AI Discriminator and~\tool{}}
    % \resizebox{0.90\linewidth}{!}{
    \begin{tabular}{c|c|c}
    \hline
        AI Discriminator &  \makecell{AI Discriminator\\ \tool{}} & \tool{} \\
    \hline
         9               &    546                    & 820 \\
    \hline
    \end{tabular}
    \label{tab:venn}
\end{table}

Table~\ref{tab:rq2_deter_aiarison} presents the experimental results of AI App identification, comparing the performance of AI Discriminator and~\tool{}. 
Among the total of 3,819 experimental Android Apps, 603 of them are classified by AI Discriminator as AI Apps, while 1,366 are recognized as AI Apps by~\tool{}, which is 242\% more than AI Discriminator. 
% To illustrate the relationships between AI Discriminator and \tool{} outcomes, we show the Venn diagram concerning the identified Android apps in Fig.~\ref{fig:venn_diagram}. 
Table~\ref{tab:venn} shows the number of AI Apps identified by AI Discriminator and~\tool{}. 
% It clearly illustrates the overlap between AI Discriminator and~\tool{} in AI Apps identification. 
% It clearly demonstrates that 546 Android apps, presenting 90.55\% $(546/603)$ and 39.97\% $(546/1366)$ in AI Discriminator and~\tool{}, respectively, are both identified as AI Apps. 
It clearly demonstrates that 546 Android Apps, accounting for 90.55\% $(546/603)$ of those identified by AI Discriminator and 39.97\% $(546/1366)$ of those confirmed by~\tool{}, are commonly recognized as AI Apps by both approaches. 
To manually inspect the nine AI apps identified solely by AI Discriminator, multiple authors downloaded the APK files from AndroZoo and dissected them with Jadx~\cite{jadx2023}. % the famous Dex to Java decompiler~\cite{jadx2023}. 
% By investigating the implementation details of the decompiled files, we find that 
Two apps are recognized as AI apps because only one special~\texttt{tflite} file is being detected, while the remaining seven apps
% ~\footnote{\eg{}, App with the sha256 value: \\EC45FCB6288EDE61077FF80885F1A8EA99600E26B37E7DCABA7069C0134A4289} 
are classified as AI apps because of keywords, such as ``DecisionTreeClassifier'' and ``KNeighborsClassifier'' in their descriptions of \texttt{json} files.
However, the description contents in the JSON files may not imply the implementation of AI capability and therefore should not be counted as convincing AI evidence.
%\terry{Why not convincing? We have similar stuff in our extracted candidates right? better to say \tool{} does not take ``descriptions'' into account, based on our definition of ``candidates''.}

To verify the 820 apps only classified by~\tool{}, we randomly sample 86 apps (with a confidence level of 95\% and a margin of error of 10\%) and cross-validated by multiple authors with the decompiler tool, Jadx~\cite{jadx2023}. 
% Consensuses will be reached by meetings if classification conflicts happen between two authors. 
% During cross-validation, we observed that some Android apps provide AI capabilities via OpenAI APIs, which cannot be detected by AI Discriminator due to its limited rule set. 
During cross-validation, we observe that OpenAI APIs are sometimes directly invoked within Java code, whereas other developers implement these API invocations via native libraries (\ie{},\texttt{.so} files).
This API invocation implementation of the cloud AI capabilities could not be detected by the AI discriminator, given its limited rule set. 
%, where HTTPS request\footnote{\url{https://api.openai.com/v1/chat/completions}} are reflected in the extracted compiled Java class file or native library~\texttt{.so} file. 
% Users of these apps need to provide an OpenAI key to explore the AI capabilities of these apps. 
In the end, we confirm that 83 of the total of 86 sampled apps are classified by~\tool{} as AI apps, showcasing a high level of identification precision. 

\begin{tcolorbox}[before skip=0.4cm, after skip=0.6cm, title=\textbf{RQ2 Findings}, left=2pt, right=2pt,top=2pt,bottom=2pt]
High precision (97.32\%) and recall (91.01\%), and the high inter-annotator reliability about the identification of AI components reconfirm the effectiveness of~\tool{} for in-wild apps. % in the wild. 
Furthermore, the experimental result of detecting more than twice as many AI apps as the rule-based approach instantiates the outperformance of~\tool{} against the SOTA one. 
\end{tcolorbox}

\subsection{\textbf{RQ3:} How is the usefulness of the AI service summary generated by~\tool{}?}

\begin{figure} [!ht]
    % \james{maybe we need to redraw this figure a bit. in the abstract and introduction, we mentioned that we employed program analysis and LLMs for the four tasks, but we didn't see them in this Figure.}
    \centering
    \includegraphics[width=\linewidth]{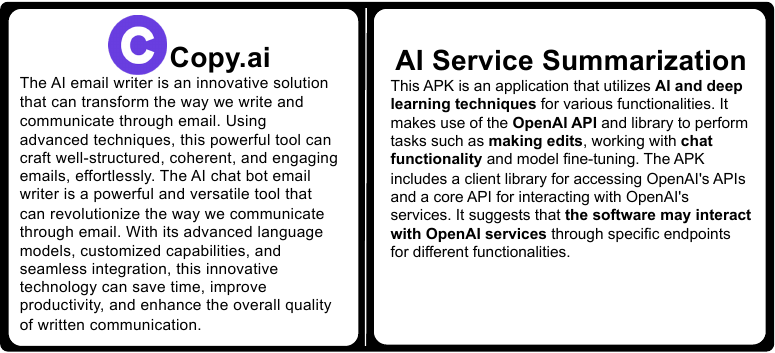}
    \caption{Description of Copy.ai (the left) and its AI service summarization (the right) generated by our framework.} 
    \label{fig:example_v3}
\end{figure}

In this research question, we seek to evaluate the usefulness of the AI summary, which is generated in Task 4 of~\tool{}. 
We conducted a user study by recruiting six developers, each with at least three years of experience in Android app analysis. 
Specifically, we randomly selected 20 Android apps identified as AI apps in RQ2, and then generated their AI summaries.
For each app, we paired each \tool{} generated AI summary with its corresponding app description retrieved from the Google Play store via the google-play-scraper~\cite{playscraper:2023}, and displayed them side-by-side to participants, as shown in Fig.~\ref{fig:example_v3}. 
We divided the six participants as two groups. Each group evaluated ten Android apps and each app was evaluated by three participants individually. 
Before conducting the user study, participants were instructed to download the apps and get familiar with their major features.
We then asked participants to read and compare the app description and the AI summary for each app, evaluating both from five aspects: \textit{Accuracy} (\ie{}, Do you think that the summary is accurate concerning the AI functions the app provides?), \textit{Completeness} (\ie{}, Do you think that the summary is complete about the provided AI tasks?), \textit{Consistency} (\ie{}, Do you think that the summary is concrete enough concerning the AI functions?), and \textit{Recommendations for app users} (\ie{}, Would you like to recommend the summary to other app users?) and developers (\ie{}, Would you like to recommend the summary to other app developers to inspire new app development?). Each dimension was assessed using a 10-point Likert scale, with scores ranging from 1 (indicating poor quality) to 10 (indicating high quality).
% For every selected app, \tool{} was used to generate an app description summary via Task 4 (AI services summarization) by feeding the identified AI components and their analyzed corresponding AI capability description. % in \tool{} and provide the generated summary and the determined AI candidates to participants. 
% by answering 5 linear scale questions with the rating scale spanning from 1 to 10. Lower rated scores indicate worse quality. 

%And also please refer to our publication site~\cite{anonyzenodo} for detailed question designs and responses. 

\begin{table*}[!ht]
    \centering
    \caption{Average and median scale values voted for sampled Android apps by six participants. ``Recom.'' stands for Recommendation, ``devs'' stands for developers, ``Med.'' stands for median.}
    \begin{threeparttable}
    \resizebox{0.98\linewidth}{!}{
    \begin{tabular}{c|c|c|c|c|c|c|c|c|c|c}
    \hline
      \multicolumn{1}{c|}{\multirow{2}{*}{Summary Type}} &  \multicolumn{2}{c|}{Accuracy}  &   \multicolumn{2}{c|}{Completeness}  &   \multicolumn{2}{c|}{Consistency}  &  \multicolumn{2}{c|}{Recom. app users}  &   \multicolumn{2}{c}{Recom. app devs.} \\
    \cline{2-11}
         &  Mean  & Med. & Mean  & Med. &  Mean  & Med. &  Mean  & Med. &  Mean  & Med. \\
    \hline
      \tool{}      & 8.67$\pm$0.14 & 9 & 8.63$\pm$0.13 & 9 & 8.55$\pm$0.13 & 9 & 7.50$\pm$0.31 & 8 & 8.57$\pm$0.15 & 9 \\
    \hline
      App Description & 6.35$\pm$0.45 & 8 & 6.17$\pm$0.45 & 8 & 6.02$\pm$0.45 & 8 & 6.87$\pm$0.46 & 8 & 6.47$\pm$0.45 & 8 \\
    \hline
    \end{tabular}
    }
    \end{threeparttable}
    \label{tab:humanstudy}
\end{table*}

Table~\ref{tab:humanstudy} shows the user study results measured by the mean and median across five dimensions. 
It shows that the AI summary generated by~\tool{} consistently outperformed the original app descriptions, underlining by higher mean scores (+2) and median scores (+1).
In terms of Accuracy and Completeness, the AI summary attained mean scores of 8.67 and 8.63, respectively. These results underscore the effectiveness and robustness of the proposed AI detection framework in capturing and conveying relevant AI functionalities for mobile apps.
Moreover, the standard error of the mean of the summary generated by~\tool{} is approximately one-third of the original description of the app, indicating a greater consensus among participants on the quality of the summaries generated by~\tool{}.
% The Table clearly shows that the \tool{}-generated summary is conspicuously better than the description provided by their corresponding developers. 
However, the voted scores for both the summary generated by~\tool{} and the original app description are quite close with respect to the metric of \textit{Recommendation for app users}. 
Specifically, the median value is the same, and the difference between the average scores is only 0.63, showcasing that different participants may have different preferences when recommending summaries to app users. 
%App users may not care about accuracy, completeness, and consistency as much as app developers and researchers. 
In general, the generated summaries by~\tool{} exhibiting higher accuracy, completeness, and consistency are well-received by developers and researchers; however, they may be less appreciated by downstream app users. 
% However, some participants believe that app users would gain benefits from such descriptions as well. 
To gain more valuable insights from app users' perspective, we intend to expand the user study to common app users as future work. 

\begin{tcolorbox}[before skip=0.4cm, after skip=0.6cm, title=\textbf{RQ3 Findings}, left=2pt, right=2pt,top=2pt,bottom=2pt]
The user study shows that the AI service summary generated by~\tool{} is constantly preferred by practitioners in terms of accuracy, completeness, consistency, and recommendation.
\end{tcolorbox}

\subsection{\textbf{RQ4:} To what extent can \tool{} uncover prevalent AI capabilities and usage patterns in real-world Android Apps?}
% \jiawei{In this part we add some insights gained from users' reviews of the apps to see the real usage of AI in pupular Android apps.}
% Regarding this research question, we focus on the identified AI components and characterize the potential services they could provide. 
\begin{table}[!ht]
    \centering
    \caption{Different types of identified AI components.} %\terry{You can also check on \% apps use APIs, \% apps use Models,  \% https request}}
    \begin{threeparttable}
    % \resizebox{0.94\linewidth}{!}{
    \begin{tabular}{c|c|c|c|c||c}
    \hline
        \makecell{Data \\ Type} &  \makecell{Package \\ API}  &  Model  &  \makecell{HTTP \\ Requests}  &  Others  & Total \\
    \hline
        \#. items  &       10,285     &   644   &   120       &   32     &  11,081 \\
    \hline
    \end{tabular}
    % }
    \end{threeparttable}
    \label{tab:data_type}
\end{table}

\begin{figure}[!ht]
    \centering
    \includegraphics[width=0.96\linewidth]{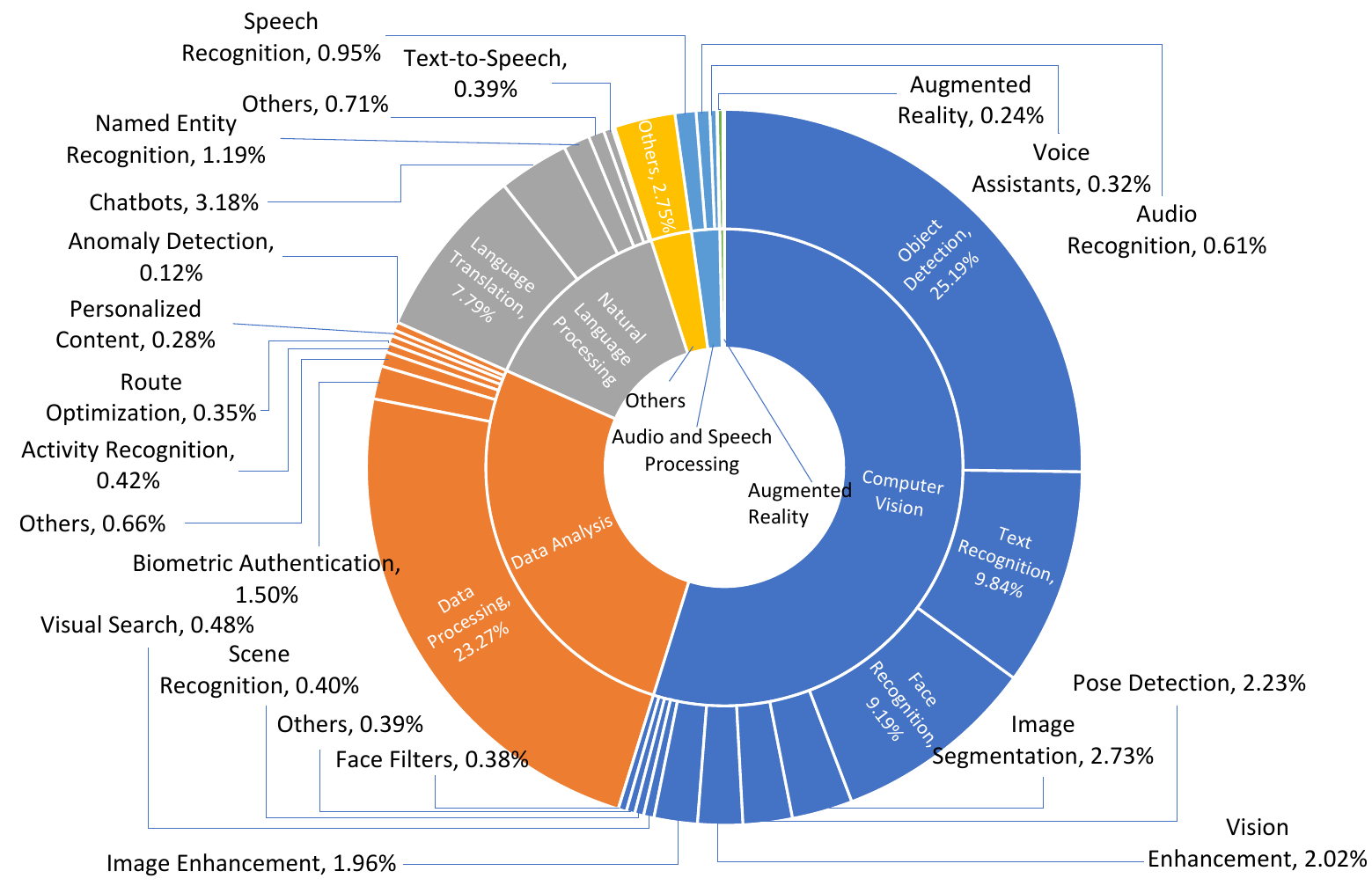}
    \caption{Detailed tasks and corresponding AI domains of the AI components.}
    \label{fig:domain_task}
\end{figure}

Regarding this research question, we seek to classify the specific domains and tasks of the identified AI components in the determined AI apps in RQ2. 
Table~\ref{tab:data_type} shows the number of AI components in different categories. 
The majority of the AI components (92.82\%, 10285/11081) lie within packages and their corresponding APIs. 
The second largest category of AI components is model, which refers to on-device AI model files, such as~\texttt{.tflite} and~\texttt{.caffemodel}. 
Apart from package names, APIs, and on-device AI models, we observe that some apps rely on HTTP requests (\eg{}, OpenAI APIs) to access cloud-based AI services, which has rarely been discussed in previous studies~\cite{xu2019first,li2022ai}. 
The remaining components exhibiting limited shared patterns are collectively classified under the category of \textit{Others}. 

\begin{table}[!ht]
    \centering
    \caption{Number of Android apps in different domains. $DA$: Data Analysis; $CV$: Computer Vision; $NLP$: Natural Language Processing; $ASP$: Audio and Speech Processing; $AR$: Augmented Reality.}
    \begin{threeparttable}
    % \resizebox{0.96\linewidth}{!}{
    \begin{tabular}{c|c|c|c|c|c||c}
    \hline
       Domains   &  DA  & CV   &  NLP  &  ASP   &  AR &  Others \\
    \hline
       \#. Apps  &  643   &  634  &  43  &  30 &  2 &  14  \\
    \hline
    \end{tabular}
    % }
    \end{threeparttable}
    \label{tab:app_domain}
\end{table}

With these identified AI components, we first reviewed the analysis results to design a classification scheme that captures a wide range of AI domains and tasks. 
% we quickly skim the analysis results and define a classification scheme that covers diverse domains and tasks. % and domains. 
% However, since there are a significant number of AI candidates, it is impractical to perform the classification manually. 
Due to the large number of AI components, we employ ChatGPT to automatically categorize each AI component into a corresponding domain and task.
% Additionally, ChatGPT is employed to automatically classify the AI domains and tasks for each AI component due to the substantial number of AI components. 
Fig.~\ref{fig:domain_task} presents the distributions of classified domains and tasks of the determined AI components. 
The area of each colored sector represents its corresponding proportion. 
At the top level, AI components are classified into six common AI domains, including Computer Vision (54.80\%), Data Analysis (26.85\%), Natural Language Processing (13.38\%), Others (2.75\%), Audio and Speech Processing (1.90\%), and Augmented Reality (0.32\%). The category~\textit{Others} contains AI components that are difficult to classify clearly. 
%, which is listed in the order of its proportion among the total number of AI candidates. 
% We note that some candidates are classified as \textit{Others} where the AI candidates are more miscellaneous.
The results show that the dominant AI capabilities are still oriented toward Computer Vision and most AI components target the tasks of Object Detection (45.97\%), Text Recognition (17.95\%), and Face Recognition (16.77\%).
The second largest AI domain is Data Analysis, which is occupied mostly by the task of Data Processing (86.69\%), such as tensor operations and input/output data parsing. 
%The remaining two domains covered by AI candidates are Augmented Reality and Audio and Speech Processing, which is . 

Based on classified AI components, we further categorize the domains and tasks at the app level. 
For each application, we begin by tallying the occurrence of domains and tasks associated with its AI components, and then categorize the app under the domain and task with the highest frequency. 
Table~\ref{tab:app_domain} shows the number of apps in different domains. 
The majority of AI apps fall into the Data Analysis domain, whereas most AI components are categorized under Computer Vision. 
This discrepancy is primarily due to the reuse of common components across various apps, with a significant concentration in the Computer Vision category. 
In addition, the number of apps in both the Computer Vision and Data Analysis domains is nearly equal, making them the two most prominent categories in both AI component classification and app-level categorization. 

In summary, the majority of AI components predominantly utilize machine learning and data processing algorithms to offer vision-related features, aiding in everyday tasks like face recognition and handwritten character recognition. 
In contrast, virtual reality functions are minimally incorporated in these apps. 
% This distribution likely stems from long-standing research and user demand for machine learning and vision-related functions. 
This could be due to the fact that virtual reality functions may lack compelling application scenarios and are not currently considered essential by developers and users. 

\begin{tcolorbox}[before skip=0.4cm, after skip=0.6cm, title=\textbf{RQ4 Findings}, left=2pt, right=2pt,top=2pt,bottom=2pt]
%To empower AI capabilities, app developers incorporate all sorts of basic AI abilities via third-party library integration reflected by so many packages import and APIs invocations. 
The most widely used AI services among Android apps and their underlying AI components are Computer Vision (54.80\%) and Data Processing (26.85\%), respectively. 
This may be attributed to the widespread user demands for vision-related functionalities, which are commonly supported across various types of mobile apps. 
% This may result from popular daily tasks to fulfill for users via all kinds of mobile apps. 
In contrast, the augmented reality oriented ones are the least provided due to the limited number of compelling real-world application scenarios, despite the concept’s novelty and growing popularity. 
%Even though Android apps are implemented on different cross-platform frameworks,~\tool{} could still be valid in AI capability identification. 
\end{tcolorbox}

\section{Discussion}
We now discuss the inherent limitations (Section~~\ref{subsec:limit}) and possible implications for both researchers and practitioners based on the outcomes of our study (Section~\ref{subsec:implication}). 

%, and also present some potential avenues for promising future research that could be

\subsection{Limitations}
\label{subsec:limit}
\subsubsection{LLM Hallucination}
Hallucination, referring to incorrect or fabricated responses to user inquiries, and inconsistency in responding to identical queries are inherent limitations of LLMs. 
In this study, we inquire about ChatGPT with our extracted implementation details (\ie{}, components) of Android apps to determine whether they are AI-related or not. 
However, the responses given by ChatGPT may not be as accurate as they really are due to hallucinations and output inconsistency. 
To mitigate such threats, we employ carefully crafted prompts and leverage large-scale experimental datasets to reduce the likelihood of hallucinated outputs and enhance response consistency. Our manual inspection of the sampled experimental results indicates that \tool{} delivers reliable and accurate analyses, demonstrating its robustness despite the inherent limitations of LLMs.

\begin{figure}[!ht]
    \centering
    \includegraphics[width=0.75\linewidth]{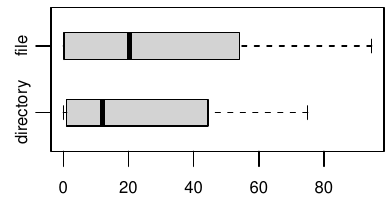}
    \caption{The percentage of obfuscated Java file names and their directory names.}
    \label{fig:obfuscation}
\end{figure}
\subsubsection{Obfuscations in Android apps}
% Obfuscation techniques, such as string encryption and Java reflection, are generally utilized to protect software artifacts against reverse engineering to keep vital intellectual property safe for its owners. 
% Android obfuscation is in fact pervasive as revealed by existing research~\cite{dong2018understanding,glanz2020hidden}. 
% % In this study, we reversed Android apps to extract the implementation details, including APIs, HTTP requests in Java files, and dynamic libraries. 
% In this study, Android apps were reverse-engineered to extract meaningful implementation details, such as APIs and HTTP requests embedded in the source code. 
Obfuscation techniques, such as string encryption (\eg{}, the obfuscated API: $<$d.f.e.x.b: void $<$init$>$ (int)$>$) and Java reflection, are widely adopted in Android apps to safeguard intellectual property against reverse engineering, as extensively evidenced by existing research~\cite{dong2018understanding,glanz2020hidden}.
% The obfuscation in Android apps, therefore, hinders the valid information extraction in our study (\eg{}, the obfuscated API: $<$d.f.e.x.b: void $<$init$>$ (int)$>$) and consequently undermines the validity of our analysis. 
% It is a long-standing research obstacle in the research area of reverse engineering. 
% The existing literature has proposed approaches to deobfuscate the obfuscated elements in Android Apps, such as APIs and constant strings. 
We have attempted to replicate some deobfuscation artifacts, such as StringHound~\cite{glanz2020hidden} and simplify~\cite{guillot2010automatic}, to recover the obfuscated items in the Android apps in our study. 
However, these tools either failed to execute successfully in our environment or produced unsatisfactory results.
To evaluate how significantly obfuscations affect the final results, we measured the ratio of obfuscation in Android apps in \textit{Dataset1}. 
Fig.~\ref{fig:obfuscation} shows the percentage of obfuscated Java file names and the directory names they reside in. 
The median value of the obfuscated Java file names and their corresponding folder names account for 11.94\% and 20.23\%, respectively. 
While obfuscation may inevitably introduce some limitations, the substantial portion of unobfuscated implementation details still enables us to extract meaningful insights and supports the reliability of our overall analysis and summarization.
% Although obfuscation could inevitably compromise our research conclusion to some extent, the remaining majority of unobfuscated implementation details could still provide us with valid, meaningful information and bolster our analysis summarization. 

\subsection{Implications}
\label{subsec:implication}

\subsubsection{Security mitigation for software artifacts.} 
% Security patches are constantly merged into vulnerable software artifacts, such as the Android Operating System and the app involved third-party libraries, to prevent severe software compromising problems (\eg{}, confidential information leaking) from happening. 
Our study reveals that by asking LLM about the implementation basics, it could respond with accurate, at least to some extent, descriptions of whether they are AI-related or not. 
% However, LLM can but is not limited to only responding to classification inquiries. 
% With careful and heuristic prompt design, LLM can also be used to do code generation, bug fixes, etc. 
One possible promising future research direction would be to use LLM as a neural knowledge to inquire if the API involves some existing vulnerabilities, for example, the vulnerability record in public Common Vulnerabilities and Exposures (CVE). 
Both researchers and developers could benefit from LLM-based vulnerability inquiries in that they could efficiently and effectively respond to potential vulnerability fixes throughout the software engineering process. 
Consequently, more reliable and secure artifacts could be achieved. 

\subsubsection{Software artifacts analysis for researchers.}
Our research on the summarization of software functionality in RQ3 validates the potential of LLM in code interpretation and summarization, specifically within the context of Java as a single programming language. 
% However, what we do in this study focuses on Java code.
The widespread use of cross-platform frameworks necessitates code interpretation and summarization across multiple programming languages, as these frameworks often integrate components written in different languages to ensure compatibility across platforms.
However, it is not straightforward to merge the interpretation in different programming languages due to the interactions between them. 
Detailed analysis of interactions between different languages and careful design of prompts are necessary to make full use of the powerful potential offered by LLMs, which also paves a promising research direction for our research communities. 

\subsubsection{Advanced AI identification approaches \& questionnaire with app developers.}
% In the AI app identification process, our manual analyses reveal that it is hard for us to accurately identify any AI candidates for some self-claimed AI Android apps. 
% More specifically, we found that some Android apps claim that they are implemented with advanced AI functionalities and tasks in their app descriptions provided on the Google Play Store. 
% However, we could not detect any AI candidates by dissecting the experimental APKs, even by manual analysis. 
% In the AI app identification process, our manual analysis reveals that some self-claimed AI Android apps, despite describing advanced AI functionalities in their Google Play Store descriptions, contain no detectable AI components upon APK dissection, even with manual inspection.
% The one possible reason for this is that we may not capture all implementation detail as the APK is now released with the new technique App bundle~\cite{appbundle:2023}, which would reduce the size of the installed APK on customers' devices and only download and install the necessary parts as needed. 
In the AI app identification process, we found that some self-claimed AI Android apps, despite describing advanced AI features in their Play Store listings, contained no detectable AI components in the released APKs. 
A possible reason is the use of Android App Bundles~\cite{appbundle:2023}, which deliver only essential components at the installation time and provide the necessary AI functions in the on-demand parts, preventing full inspection of the implementation details. 
% The app developers could implement the necessary AI functions in the on-demand parts. 
To enable comprehensive analysis, it is imperative to adopt a new approach capable of harvesting all parts of an Android app, not only the basic ones but also those loaded on demand. 
Moreover, not all AI function implementations in Android apps are trivial. 
There may be some non-trivial implementations that are transparent to us. 
Therefore, a questionnaire with these popular Android app developers would provide us with a few great insights for further app analysis.

\section{Related Work}
\subsection{AI Mobile Apps Study}
Xu~\etal{}~\cite{xu2019first} first proposed an APK analysis tool,~\texttt{DL Sniffer} to empirically investigate the use of deep learning techniques in Android apps. 
The DL Sniffer dissects the released APK and identifies the utilization of 16 popular deep learning frameworks via two separate modules. 
% The first one recognizes specific strings (\eg{}, ``TF\_AllocateTensor'') in the section~\textit{rodata} of native shared libraries (\ie{}, file ends with ``.so'') via some powerful Linux command~\cite{readelf:2023,objdump:2023}. 
% The other one identifies AI special APIs in DEX binary files by de-compiling the binary file into smali code via~\cite{dex2jar:2023}. 
Experiments on the large-scale dataset containing 16,500 Android apps reveal that top apps are AI techniques early adopters, especially photo beauty apps, and the on-device AI is actually lighter than practitioners expect. 
In addition, the DL framework developers should take more time to optimize and protect frameworks. 
% Sun~\etal{}~\cite{sun2021mind} focused on the security part of DL frameworks in Android apps and curated an even larger dataset contacting 46,753 Android apps from the US and Chinese app markets. 
% a static DL framework analysis tool, 
Sun~\etal{}~\cite{sun2021mind} proposed ModelXRay to determine whether on-device AI frameworks are adopted and to check if the adopted framework is protected or encrypted in a large dataset containing 46,753 Android apps. 
Their experimental results show that 41\% Android apps do not protect the adopted DL models, and even some of the protected models still have a weak status, which would result in the leakage of private information and economic losses. 
Li~\etal{}~\cite{li2022ai} conducted an exploratory study and released a tool, AI Discriminator, to analyze the adoption of AI technologies in Android apps. 
They first refined the rules used in the identification of AI app and successfully recognized 56,682 AI apps from a total of 7,259,232 mobile apps provided on AndroZoo~\cite{allix2016androzoo}. 
% With the historical versions of apps considered, 
They concluded that the AI models utilized in Android apps are not updated regularly and are not protected carefully. 
% What's more, AI apps always request access to various types of sensitive user data. 
In addition, they analyzed reviews for the determined AI apps and revealed that users always express positive attitudes towards the adoption of advanced AI technologies. 

\subsection{LLMs in Software Engineering}
Feng~\etal{}~\cite{feng2023prompting} tried to reproduce Android bugs with the help of LLMs. 
They proposed a lightweight bug replay tool, AdbGPT, to automatically rerun Android APKs and trigger bugs according to collected bug reports. 
%The AdbGPT first utilizes prompt engineering
%, working as real app developers, 
%to extract the necessary steps for bug replay from the given bug reports. 
%And then prompts LLMs to run Android apps under the guidance extracted previously. 
Their extensive experiments reveal that AdbGPT could 
% obtain a higher accuracy in necessary step extraction compared with the state-of-the-art traditional approaches and finally 
successfully reproduce 81.3\% bugs, which is much better than baselines. 
Their approach demonstrates the effectiveness of LLMs in assisting traditional software research problems.
Tian~\etal{}~\cite{tian2023chatgpt} focused on the research tasks of code generation, program repair, and code summarization and conducted an exploratory study to investigate performance in the aforementioned practices. 
%They curated datasets from LeetCode 2016-2020 and LeetCode 2022, and construct prompts to generate code with problem descriptions, repair buggy code implementations, and summarize code intentions. 
The experimental results not only reveal that ChatGPT is effective in common software practices but also unveil that limitations exist in LLMs, such as long and detailed prompt descriptions that would hinder the effectiveness of ChatGPT. 
Ma~\etal{}~\cite{ma2023scope} conducted a study to touch on the limitations of the interpretability of ChatGPT in three types of software engineering jobs, including understanding syntax, static behaviors, and dynamic behaviors.
%, in which abstract syntax tree (AST), control flow graphs (CFG), and call graphs (CG) are leveraged to measure the interpretability. 
% including C, Java, Python, and Solidity, 
Performing ChatGPT on different programming languages, the authors concluded that ChatGPT does have the ability to construct AST but it still produces incorrect or fake facts with respect to dynamic semantic understanding. 
%It is necessary for researchers to propose effective methods to verify the correctness of the generated outputs from ChatGPT.
Li~\etal{}~\cite{li2023hitchhiker} carried out an LLM-assisted static analysis with respect to use-before-initialization (UBI) bugs for C/C++ programs. 
%They proposed an automatic agent that interacts with both a static analysis tool and an LLM. 
% With the state-of-the-art static analysis tool and meticulous design of prompts, the authors could successfully overcome several challenges, such as modeling bugs specific to the problem, which are hard to manage within separate research directions. 
They conducted experiments on real potential UBI bugs, achieved higher precision and recall, and even identified 13 previously untouched ones, revealing the effectiveness of LLM in assisting software engineering problems.

\section{Conclusion}
In this paper, we have proposed an automated approach~\tool{}, which employs program analysis techniques to extract detailed implementation components from Android apps and utilizes a large language model (ChatGPT in this study) to interpret and summarize the underlying AI services. 
Specifically, our approach addresses the intuitive questions, including ``Does the application incorporate AI capabilities'' and ``What specific types of AI functionalities are embedded''. 
% We conduct experiments on both the existing approach determined AI apps and the self-curated in-wild apps. 
The experimental results on both AI apps previously identified by a rule-based approach and in-wild apps demonstrate that~\tool{} could achieve a precision of 98.05\% and a recall of 93.31\% on the former AI apps and a precision of 97.32\% and a recall of 91.01\% on the latter in-wild apps regarding AI components identification. %for the identification of AI components. 
Moreover,~\tool{} not only identifies more than twice as many AI apps in the wild compared to the SOTA rule-based approach, but also provides meaningful functional descriptions. 
To further evaluate its practical utility, we have also conducted a user study on the usefulness of~\tool{} and the study results confirm that the app summary generated by~\tool{} is much preferred by participants with respect to the metrics of accuracy, completeness, consistency, and recommendation. %  for app developers
% We then define a set of tasks and domains and categorize the identified AI candidates and determined AI apps to the manually classified tasks and domains. 
%Apart from AI candidates, we also classified the identified AI apps to the corresponding domains according to the largest domain proportion of their integral candidates. 
Additionally, our analyses reveal that the most prevalent AI services are related to machine learning (26.85\%) and computer vision (54.80\%), whereas virtual reality based functions are actually the least presented, despite their growing popularity and increasing interest among both AI researchers and early adopters. 
Furthermore, with these findings, we outline several promising research directions to inform and inspire future work in this area. 
%even though it is an increasingly popular concept and gains growing interest from both AI researchers and early adopters nowadays. 
% Furthermore, with our findings, we also provide a few promising research insights for our fellow researchers. 
% \bibliographystyle{IEEEtran}

% %\section{Data Availability}
% \textbf{Data Availability:}
% The source code and datasets are all made publicly available in our artifact package via the following link:
% %\begin{center}
% \url{https://zenodo.org/records/15522834}
%\end{center}

\section{Declarations}

\textbf{}

\textbf{Funding:} Not applicable.

\textbf{Ethical approval:} Not applicable.

\textbf{Informed consent:} Not applicable.

\textbf{Author Contributions:} Pei Liu led the project, oversaw its administration, and drafted the original manuscript. Terry Yue Zhuo developed and implemented the framework. Jiawei Deng curated the dataset and performed the experimental analyses. Zhenchang Xing, Qinghua Lu, Xiaoning Du conceived and designed the study and contributed to manuscript writing. Hongyu Zhang provided critical revisions and editorial improvements.

\textbf{Data Availability Statements:} The source code and datasets are all made publicly available in our artifact package via the following link: \url{https://zenodo.org/records/15958951}

\textbf{Conflict of Interest:} The authors declare that they have no known competing financial interests or personal relationships that could have appeared to influence the work reported in this paper.

\textbf{Clinical trial number:} Not applicable.

\bibliographystyle{spmpsci}      % mathematics and physical sciences
\bibliography{mainref}

@String{Computing = "Computing" }

@String{Computer = "{IEEE} Computer" }

@String{Springer = "Springer-Verlag" }

@manual{readelf:2023,
  author = {readelf},
  title  = {readelf - display information about ELF files},
  note    = {\url{https://man7.org/linux/man-pages/man1/readelf.1.html}}
}

@manual{objdump:2023,
  author = {objdump},
  title  = {objdump - display information from object files},
  note    = {\url{https://man7.org/linux/man-pages/man1/objdump.1.html}}
}

@manual{dalvik:2022,
  author = {Dalvik},
  title  = {Dalvik bytecode},
  note    = {\url{https://source.android.com/docs/core/runtime/dalvik-bytecode}},
}

@article{wei2022chain,
  title={Chain-of-thought prompting elicits reasoning in large language models},
  author={Wei, Jason and Wang, Xuezhi and Schuurmans, Dale and Bosma, Maarten and Xia, Fei and Chi, Ed and Le, Quoc V and Zhou, Denny and others},
  journal={Advances in Neural Information Processing Systems},
  volume={35},
  pages={24824--24837},
  year={2022}
}

@inproceedings{guo2018cloud,
  title={Cloud-based or on-device: An empirical study of mobile deep inference},
  author={Guo, Tian},
  booktitle={2018 IEEE International Conference on Cloud Engineering (IC2E)},
  pages={184--190},
  year={2018},
  organization={IEEE}
}

@article{hu2023first,
  title={A First Look at On-device Models in iOS Apps},
  author={Hu, Han and Huang, Yujin and Chen, Qiuyuan and Zhuo, Terry Yue and Chen, Chunyang},
  journal={ACM Transactions on Software Engineering and Methodology},
  volume={33},
  number={1},
  pages={1--30},
  year={2023},
  publisher={ACM New York, NY}
}

@manual{appbundle:2023,
  author = {Google for Developers},
  title  = {About Android App Bundles},
  note    = {\url{https://developer.android.com/guide/app-bundle}},
  year   = {2024}
}

@manual{stable:2023,
  author = {Stable Diffusion},
  title  = {Stable Diffusion Online},
  note    = {\url{https://stablediffusionweb.com/}},
}

@inproceedings{abadi2016tensorflow,
  title={$\{$TensorFlow$\}$: a system for $\{$Large-Scale$\}$ machine learning},
  author={Abadi, Mart{\'\i}n and Barham, Paul and Chen, Jianmin and Chen, Zhifeng and Davis, Andy and Dean, Jeffrey and Devin, Matthieu and Ghemawat, Sanjay and Irving, Geoffrey and Isard, Michael and others},
  booktitle={12th USENIX symposium on operating systems design and implementation (OSDI 16)},
  pages={265--283},
  year={2016}
}

@manual{caffe2:2023,
  author = {Caffe2},
  title  = {Caffe2 is a lightweight, modular, and scalable deep learning framework. Building on the original Caffe, Caffe2 is designed with expression, speed, and modularity in mind.},
  note    = {\url{https://github.com/pytorch/pytorch/tree/main/caffe2}}
}

@manual{coreml:2023,
  author = {Core ML},
  title  = {Integrate machine learning models into your app.},
  note    = {\url{https://developer.apple.com/documentation/coreml}}
}

@manual{ncnn:2023,
  author = {ncnn},
  title  = {ncnn is a high-performance neural network inference computing framework optimized for mobile platforms.},
  note    = {\url{https://github.com/Tencent/ncnn}}
}

@article{zhao2022survey,
  title={A survey of deep learning on mobile devices: Applications, optimizations, challenges, and research opportunities},
  author={Zhao, Tianming and Xie, Yucheng and Wang, Yan and Cheng, Jerry and Guo, Xiaonan and Hu, Bin and Chen, Yingying},
  journal={Proceedings of the IEEE},
  volume={110},
  number={3},
  pages={334--354},
  year={2022},
  publisher={IEEE}
}

@article{li2022ai,
  title={AI-driven Mobile Apps: an Explorative Study},
  author={Li, Yinghua and Dang, Xueqi and Tian, Haoye and Sun, Tiezhu and Wang, Zhijie and Ma, Lei and Klein, Jacques and Bissyande, Tegawende F},
  journal={arXiv preprint arXiv:2212.01635},
  year={2022}
}

@inproceedings{xu2019first,
  title={A first look at deep learning apps on smartphones},
  author={Xu, Mengwei and Liu, Jiawei and Liu, Yuanqiang and Lin, Felix Xiaozhu and Liu, Yunxin and Liu, Xuanzhe},
  booktitle={The World Wide Web Conference},
  pages={2125--2136},
  year={2019}
}

@article{siu2023towards,
  title={Towards Real Smart Apps: Investigating Human-AI Interactions in Smartphone On-Device AI Apps},
  author={Siu, Jason Ching Yuen and Chen, Jieshan and Huang, Yujin and Xing, Zhenchang and Chen, Chunyang},
  journal={arXiv preprint arXiv:2307.00756},
  year={2023}
}

@inproceedings{deng2019deep,
  title={Deep learning on mobile devices: a review},
  author={Deng, Yunbin},
  booktitle={Mobile Multimedia/Image Processing, Security, and Applications 2019},
  volume={10993},
  pages={52--66},
  year={2019},
  organization={SPIE}
}

@inproceedings{sun2021mind,
  title={Mind your weight (s): A large-scale study on insufficient machine learning model protection in mobile apps},
  author={Sun, Zhichuang and Sun, Ruimin and Lu, Long and Mislove, Alan},
  booktitle={30th USENIX Security Symposium (USENIX Security 21)},
  pages={1955--1972},
  year={2021}
}

@manual{chatgpt:2023,
  author = {OpenAI},
  title  = {ChatGPT: get instant answers, find creative inspiration, and learn something new.},
  note   = {\url{https://openai.com/chatgpt}},
  year   = {2024}
}

@manual{gptmodel:2023,
  author = {OpenAI},
  title  = {Models},
  note    = {\url{https://platform.openai.com/docs/models/overview}}
}

@manual{playscraper:2023,
  author = {facundoolano},
  title  = {google-play-scraper},
  note    = {\url{https://www.npmjs.com/package/google-play-scraper}},
  year   = {2023}
}

@manual{aiapp:2023,
  author = {yinghuali},
  title  = {AI-driven Mobile Apps: an Explorative Study},
  note    = {\url{https://github.com/yinghuali/AIApp}},
  year   = {2023}
}

@manual{ndk:2023,
  author = {Android NDK},
  title  = {The Android NDK is a toolset that lets you implement parts of your app in native code, using languages such as C and C++},
  note    = {\url{https://developer.android.com/ndk}}
}

@article{brown2020language,
  title={Language models are few-shot learners},
  author={Brown, Tom and Mann, Benjamin and Ryder, Nick and Subbiah, Melanie and Kaplan, Jared D and Dhariwal, Prafulla and Neelakantan, Arvind and Shyam, Pranav and Sastry, Girish and Askell, Amanda and others},
  journal={Advances in neural information processing systems},
  volume={33},
  pages={1877--1901},
  year={2020}
}

@article{feng2023prompting,
  title={Prompting Is All Your Need: Automated Android Bug Replay with Large Language Models},
  author={Feng, Sidong and Chen, Chunyang},
  journal={arXiv preprint arXiv:2306.01987},
  year={2023}
}

@article{li2023hitchhiker,
  title={The Hitchhiker's Guide to Program Analysis: A Journey with Large Language Models},
  author={Li, Haonan and Hao, Yu and Zhai, Yizhuo and Qian, Zhiyun},
  journal={arXiv preprint arXiv:2308.00245},
  year={2023}
}

@article{tian2023chatgpt,
  title={Is ChatGPT the Ultimate Programming Assistant--How far is it?},
  author={Tian, Haoye and Lu, Weiqi and Li, Tsz On and Tang, Xunzhu and Cheung, Shing-Chi and Klein, Jacques and Bissyand{\'e}, Tegawend{\'e} F},
  journal={arXiv preprint arXiv:2304.11938},
  year={2023}
}

@article{wang2020generalizing,
  title={Generalizing from a few examples: A survey on few-shot learning},
  author={Wang, Yaqing and Yao, Quanming and Kwok, James T and Ni, Lionel M},
  journal={ACM computing surveys (csur)},
  volume={53},
  number={3},
  pages={1--34},
  year={2020},
  publisher={ACM New York, NY, USA}
}

@inproceedings{allix2016androzoo,
  title={Androzoo: Collecting millions of android apps for the research community},
  author={Allix, Kevin and Bissyand{\'e}, Tegawend{\'e} F and Klein, Jacques and Le Traon, Yves},
  booktitle={Proceedings of the 13th international conference on mining software repositories},
  pages={468--471},
  year={2016}
}

@article{ma2023scope,
  title={The Scope of ChatGPT in Software Engineering: A Thorough Investigation},
  author={Ma, Wei and Liu, Shangqing and Wang, Wenhan and Hu, Qiang and Liu, Ye and Zhang, Cen and Nie, Liming and Liu, Yang},
  journal={arXiv preprint arXiv:2305.12138},
  year={2023}
}

@inproceedings{liu2021identifying,
  title={Identifying and characterizing silently-evolved methods in the android API},
  author={Liu, Pei and Li, Li and Yan, Yichun and Fazzini, Mattia and Grundy, John},
  booktitle={2021 IEEE/ACM 43rd International Conference on Software Engineering: Software Engineering in Practice (ICSE-SEIP)},
  pages={308--317},
  year={2021},
  organization={IEEE}
}

@inproceedings{bartel2012dexpler,
  title={Dexpler: converting android dalvik bytecode to jimple for static analysis with soot},
  author={Bartel, Alexandre and Klein, Jacques and Le Traon, Yves and Monperrus, Martin},
  booktitle={Proceedings of the ACM SIGPLAN International Workshop on State of the Art in Java Program analysis},
  pages={27--38},
  year={2012}
}

@inproceedings{chakraborty2022natgen,
  title={Natgen: generative pre-training by “naturalizing” source code},
  author={Chakraborty, Saikat and Ahmed, Toufique and Ding, Yangruibo and Devanbu, Premkumar T and Ray, Baishakhi},
  booktitle={Proceedings of the 30th ACM Joint European Software Engineering Conference and Symposium on the Foundations of Software Engineering},
  pages={18--30},
  year={2022}
}

@article{fried2022incoder,
  title={Incoder: A generative model for code infilling and synthesis},
  author={Fried, Daniel and Aghajanyan, Armen and Lin, Jessy and Wang, Sida and Wallace, Eric and Shi, Freda and Zhong, Ruiqi and Yih, Wen-tau and Zettlemoyer, Luke and Lewis, Mike},
  journal={arXiv preprint arXiv:2204.05999},
  year={2022}
}

@misc{nijkamp2022conversational,
  title={A conversational paradigm for program synthesis},
  author={Nijkamp, Erik and Pang, Bo and Hayashi, Hiroaki and Tu, Lifu and Wang, Huan and Zhou, Yingbo and Savarese, Silvio and Xiong, Caiming},
  journal={arXiv preprint arXiv:2203.13474},
  volume={30},
  year={2022},
  publisher={Mar}
}

@article{jiang2023impact,
  title={Impact of code language models on automated program repair},
  author={Jiang, Nan and Liu, Kevin and Lutellier, Thibaud and Tan, Lin},
  journal={arXiv preprint arXiv:2302.05020},
  year={2023}
}

@article{zhang2022repairing,
  title={Repairing bugs in python assignments using large language models},
  author={Zhang, Jialu and Cambronero, Jos{\'e} and Gulwani, Sumit and Le, Vu and Piskac, Ruzica and Soares, Gustavo and Verbruggen, Gust},
  journal={arXiv preprint arXiv:2209.14876},
  year={2022}
}

@article{xia2023conversational,
  title={Conversational automated program repair},
  author={Xia, Chunqiu Steven and Zhang, Lingming},
  journal={arXiv preprint arXiv:2301.13246},
  year={2023}
}

@inproceedings{dong2018understanding,
  title={Understanding android obfuscation techniques: A large-scale investigation in the wild},
  author={Dong, Shuaike and Li, Menghao and Diao, Wenrui and Liu, Xiangyu and Liu, Jian and Li, Zhou and Xu, Fenghao and Chen, Kai and Wang, Xiaofeng and Zhang, Kehuan},
  booktitle={Security and Privacy in Communication Networks: 14th International Conference, SecureComm 2018, Singapore, Singapore, August 8-10, 2018, Proceedings, Part I},
  pages={172--192},
  year={2018},
  organization={Springer}
}

@inproceedings{glanz2020hidden,
  title={Hidden in plain sight: Obfuscated strings threatening your privacy},
  author={Glanz, Leonid and M{\"u}ller, Patrick and Baumg{\"a}rtner, Lars and Reif, Michael and Amann, Sven and Anthonysamy, Pauline and Mezini, Mira},
  booktitle={Proceedings of the 15th ACM Asia Conference on Computer and Communications Security},
  pages={694--707},
  year={2020}
}

@misc{cohenkappa,
  title = {Inter-rater reliability},
  howpublished = {Online},
  note = {\url{https://en.wikipedia.org/wiki/Cohen\%27s_kappa}}
}

@misc{aapt:2024,
  title = {AAPT2},

  howpublished = {Online},
  note = {\url{https://developer.android.com/tools/aapt2}}
}

@misc{jadx2023,
  title = {jadx - Dex to Java decompiler},
  howpublished = {Online},
  note = {\url{https://github.com/skylot/jadx}}
}

@article{guillot2010automatic,
  title={Automatic binary deobfuscation},
  author={Guillot, Yoann and Gazet, Alexandre},
  journal={Journal in computer virology},
  volume={6},
  number={3},
  pages={261--276},
  year={2010},
  publisher={Springer}
}

@article{touvron2023llama,
  title={Llama: Open and efficient foundation language models},
  author={Touvron, Hugo and Lavril, Thibaut and Izacard, Gautier and Martinet, Xavier and Lachaux, Marie-Anne and Lacroix, Timoth{\'e}e and Rozi{\`e}re, Baptiste and Goyal, Naman and Hambro, Eric and Azhar, Faisal and others},
  journal={arXiv preprint arXiv:2302.13971},
  year={2023}
}

@article{liu2024deepseek,
  title={Deepseek-v3 technical report},
  author={Liu, Aixin and Feng, Bei and Xue, Bing and Wang, Bingxuan and Wu, Bochao and Lu, Chengda and Zhao, Chenggang and Deng, Chengqi and Zhang, Chenyu and Ruan, Chong and others},
  journal={arXiv preprint arXiv:2412.19437},
  year={2024}
}

@article{hagendorff2023human,
  title={Human-like intuitive behavior and reasoning biases emerged in large language models but disappeared in ChatGPT},
  author={Hagendorff, Thilo and Fabi, Sarah and Kosinski, Michal},
  journal={Nature Computational Science},
  volume={3},
  number={10},
  pages={833--838},
  year={2023},
  publisher={Nature Publishing Group US New York}
}

@article{ning2024can,
  title={Can LLMs learn by teaching for better reasoning? A preliminary study},
  author={Ning, Xuefei and Wang, Zifu and Li, Shiyao and Lin, Zinan and Yao, Peiran and Fu, Tianyu and Blaschko, Matthew and Dai, Guohao and Yang, Huazhong and Wang, Yu},
  journal={Advances in Neural Information Processing Systems},
  volume={37},
  pages={71188--71239},
  year={2024}
}

@article{pan2024large,
  title={A Large-scale Investigation of Semantically Incompatible APIs behind Compatibility Issues in Android Apps},
  author={Pan, Shidong and Guo, Tianchen and Zhang, Lihong and Liu, Pei and Xing, Zhenchang and Sun, Xiaoyu},
  journal={arXiv preprint arXiv:2406.17431},
  year={2024}
}

@article{liao20243,
  title={A 3-CodGen: A Repository-Level Code Generation Framework for Code Reuse with Local-Aware, Global-Aware, and Third-Party-Library-Aware},
  author={Liao, Dianshu and Pan, Shidong and Sun, Xiaoyu and Ren, Xiaoxue and Huang, Qing and Xing, Zhenchang and Jin, Huan and Li, Qinying},
  journal={IEEE Transactions on Software Engineering},
  year={2024},
  publisher={IEEE}
}

@article{han2024chase,
  title={Do Chase Your Tail! Missing Key Aspects Augmentation in Textual Vulnerability Descriptions of Long-tail Software through Feature Inference},
  author={Han, Linyi and Pan, Shidong and Xing, Zhenchang and Sun, Jiamou and Yitagesu, Sofonias and Zhang, Xiaowang and Feng, Zhiyong},
  journal={IEEE Transactions on Software Engineering},
  year={2024},
  publisher={IEEE}
}

@inproceedings{li2024llms,
  title={Llms for relational reasoning: How far are we?},
  author={Li, Zhiming and Cao, Yushi and Xu, Xiufeng and Jiang, Junzhe and Liu, Xu and Teo, Yon Shin and Lin, Shang-Wei and Liu, Yang},
  booktitle={Proceedings of the 1st International Workshop on Large Language Models for Code},
  pages={119--126},
  year={2024}
}
\end{document}